\documentclass[fleqn,usenatbib]{mnras}
\usepackage[percent]{overpic}
\usepackage[normalem]{ulem}

\usepackage{newtxtext,newtxmath}
\usepackage{comment}
\usepackage[dvipsnames]{xcolor}
\usepackage{subcaption}
\usepackage[T1]{fontenc}
\usepackage{ae,aecompl}
\usepackage{algorithm}
\usepackage{algorithmic}

\DeclareRobustCommand{\VAN}[3]{#2}
\let\VANthebibliography\thebibliography
\def\thebibliography{\DeclareRobustCommand{\VAN}[3]{##3}\VANthebibliography}

\usepackage{graphicx}	% Including figure files
\usepackage{newtxtext,newtxmath}

\usepackage{natbib}
\usepackage{bm}         % bold face math symbols
\usepackage{xspace}
\usepackage{xcolor}
\usepackage{comment}

\usepackage{mwe}
\usepackage{caption}
\usepackage{subcaption}
\usepackage{soul}   %highlighting

\title[AI-evolved structure formation]{Benchmarking AI-evolved cosmological structure formation}

\author[Dong et al.]{Xiaofeng Dong $^{1, 2}$\thanks{E-mail:xfdong@uchicago.edu},
Nesar Ramachandra$^{2, 3}$, 
Salman Habib$^{2, 3}$,
Katrin Heitmann$^{2}$
\\$^{1}$ Department of Physics, University of Chicago, Chicago, IL 60637, USA.
\\$^{2}$ High Energy Physics Division, Argonne National Laboratory, 9700 South Cass Avenue, Lemont, IL 60439, USA.
\\$^{3}$ Computational Science Division, Argonne National Laboratory, 9700 South Cass Avenue, Lemont, IL 60439, USA.
}

\date{Accepted XXX. Received YYY; in original form ZZZ}

\pubyear{2025}

\begin{document}

 \maketitle

\begin{abstract}
The potential of deep learning-based image-to-image translations has recently attracted significant attention. One possible application of such a framework is as a fast, approximate alternative to cosmological simulations, which would be particularly useful in various contexts, including covariance studies, investigations of systematics, and cosmological parameter inference. To investigate different aspects of learning-based cosmological mappings, we choose two approaches for generating suitable cosmological matter fields as datasets: a simple analytical prescription provided by the Zel'dovich approximation, and a numerical N-body method using the Particle-Mesh approach. The evolution of structure formation is modeled using U-Net, a widely employed convolutional image translation framework. Because of the lack of a controlled methodology, validation of these learned mappings requires multiple benchmarks beyond simple visual comparisons and summary statistics.  A comprehensive list of metrics is considered, including higher-order correlation functions, conservation laws, topological indicators, and statistical independence of density fields. We find that the U-Net approach performs well only for some of these physical metrics, and accuracy is worse at increasingly smaller scales, where the dynamic range in density is large. By introducing a custom density-weighted loss function during training, we demonstrate a significant improvement in the U-Net results at smaller scales. This study provides an example of how a family of physically motivated benchmarks can, in turn, be used to fine-tune optimization schemes -- such as the density-weighted loss used here -- to significantly enhance the accuracy of scientific machine learning approaches by focusing attention on relevant features.
\end{abstract}

\section{Introduction}
\label{sec:intro}
In the era of `precision cosmology', cosmological N-body and hydrodynamical simulations play a key role in the study of the large-scale structure of the Universe and are essential to interpreting the unprecedented amount of observational data available from sky surveys (for reviews, see~\citealt{Dolag2008, Angulo2022}).
%\citep{Heitmann2014,Habib_2016,Garrison2021,Bayer2024,Falck2021,Springel2005,Angulo2012,Klypin2016}.
 By tracking the gravitational evolution of dark matter and baryonic components, these simulations enable a rigorous interpretation of the extensive observational data collected by modern sky surveys \citep{Springel2005,Vogelsberger2014,Schaye2015,Dolag2008}. Building on the early work of \cite{klypin1983} and \cite{davis1985evolution}, numerical models not only shed light on the properties of dark matter and dark energy and the origins of primordial fluctuations, but also provide critical constraints on fundamental parameters such as the neutrino mass sum. Moreover, these simulations generate detailed synthetic survey observations that are essential for designing and optimizing observational campaigns, as well as for investigating astrophysical and instrumental systematics (e.g., \citealt{Korytov2019,2021ApJS..253...31L}). 

Cosmological simulations display significant diversity, ranging from gravity-only, large-volume simulations to smaller, more detailed and physically rich hydrodynamic simulations that target the details of galaxy formation. A small set of runs, or sometimes even a single simulation, may be enough to address the problem at hand. However, it is often the case that a large set of simulations (ensembles over parameters and realizations) is required. These can be used to generate data for covariance studies~\citep{2018arXiv180901669T} or to build emulators for precision predictions, solving inverse problems in cosmology (e.g., determining cosmological parameters from a set of observations). It is computationally challenging to evolve many billions or trillions of particles at high enough resolution, even for a single simulation. For the large number of simulations often required when running ensembles, the computing requirements can quickly become prohibitive.

The sizes and resolution requirements for ensemble campaigns also vary considerably. Applications restricted to emulation of summary statistics (e.g., density power spectrum, halo mass function) in the nonlinear regime of structure formation may require hundreds of simulations at near state-of-the-art resolution \citep{Heitmann2016, derose2018aemulus}, while covariance studies may require thousands (or many more), but at lower resolutions \citep{aemulus2025howmany}. In some cases (e.g., field reconstruction studies), it is essential not only to generate summary statistics but also to have the full simulation results available.

Given the potential resource constraints associated with running cosmological simulations, different strategies have been considered to reduce the total computational cost. These either simplify the computations, e.g., lower resolution, simplified physics models (\citealt{Angulo2021}), or reduce their number, e.g., adaptive and optimal sampling, use of scaling  \citep{Chartier2020, Wraith2009}. Yet another approach (possibly involving multi-resolution ideas) is to consider replacing the simulations entirely via a generative model based on deep learning (DL) applied to a training data set built on simulation results \citep{Mustafa_2019,Perraudin2020,Dai2020,Li2021}. The two key questions that arise are: 1) Is it technically feasible for the generative model to produce data at the required level of accuracy? 2) To reach the demanded level of performance, what is the required training cost (since a very large training cost could potentially nullify the advantage of the DL-based approach)? Our purpose here is to pursue these two questions in a simplified, but sufficiently useful setting.

Recent DL advances have highlighted the potential of the technique in capturing highly complex functions and mappings, thereby attracting attention in various scientific domains, including cosmology~\citep{Ntampaka2019, schmelzle2017cosmological, chardin2019deep, gunther2022cosmicnet}.  
Deep neural networks can serve as universal approximations, having the ability to learn underlying distributions of data and to predict a wide variety of observables, including summary statistics, as well as full sets of simulated fields (e.g., 3-d and 2-d density and velocity fields). %Deep generative modeling techniques thus introduce new potentials to build computationally cheaper emulators through density estimators built out of neural networks~\citep{Mustafa_2019}.  
Cosmological applications of convolutional neural networks and deep learning in the simulation context include the generation of weak lensing convergence maps~\citep{Mustafa_2019}, parameter inference using weak gravitational lensing~\citep{ribli2019improved}, parameter regression from data simulations, improvement of differentiating between dark energy and modified gravity cosmologies~\citep{peel2019distinguishing}, and de-noising of lensing maps~\citep{shirasaki2019denoising}. Machine learning data analysis tools have also made their impact in various cosmological contexts, such as reducing scatter in galaxy cluster mass estimates, tightening cosmological parameter constraints for weak lensing maps, extracting cosmological parameters from large-scale structure, classifying sources driving reionization, and high signal-to-noise extraction of the projected gravitational potential from cosmic microwave background maps~\citep{Ntampaka2019}.

If DL methods can successfully capture the full complexity of cosmic evolution, they can provide a valuable approach in addressing the aforementioned computational bottleneck, and perhaps even eliminate it in some cases. However, the ability of DL models to make sufficiently accurate and physically meaningful predictions is yet to be fully investigated. The problem is exacerbated by the `black box' nature of the methods and it is difficult to predict a priori what their error properties might be. Similarly, it is not obvious how well they can describe the detailed information present in cosmological simulations in ways that do not violate physical constraints. Finally, it would be important to know the sizes of the training sets needed to build a sufficiently useful DL model. The approach would not be successful if the amount of effort expended on training is similar to or exceeds that needed for a more brute force computation-based approach. Asked in another way, can DL models, without any domain knowledge of cosmology, capture all the intricacies of nonlinear gravitational clustering? And in what way can AI-based emulators best complement standard numerical approaches?

In current efforts applying DL to cosmology, more attention needs to be paid to whether 1) the results are sufficiently accurate and physically consistent, and 2) how to benchmark them by imposing a set of physically motivated metrics to test for these properties.  Methods to alleviate weaknesses of DL approaches include setting up physics-informed terms in the loss function \citep{cao2022physics} and architectural modifications implemented into the network to facilitate the correct physical boundary conditions, such as translation and rotational symmetries. Despite such efforts, there is no guarantee that the prediction results follow physical laws in the sense of being a controlled approximation to the underlying equations. For meaningful scientific applications, a variety of benchmarks and tests on the prediction/generation results need to be conducted to quantitatively assess the accuracy and/or deviation from expected behavior.  

The other impediment that deep learning methods normally encounter is data scalability. Prototypical studies normally perform well on smaller-sized simulation data, while for AI methods to be applied for real-life purposes in cosmology, the scaling with data size would require much more GPU memory for deep neural networks, and oftentimes it is computationally forbidding to train and apply such models. This problem can be appreciated by noting that large-scale cosmological simulations can have a 3-d dynamic range of roughly a million to one (i.e., the largest scales in the simulation are roughly a million times bigger than the smallest resolved scales), whereas most DL approaches studied so far cover only a dynamic range of one to three orders of magnitude.

In the work presented here, we address a number of the issues mentioned above, primarily related to assessing the fidelity of DL-based methods and some aspects related to convergence. The dynamic range considered is modest, since, as we will show, many of the issues being investigated are manifest already in this relatively simple case. We do not fully address questions of scalability for now, since they become relevant only after questions of accuracy and convergence are more completely resolved.

In our work, we model cosmic evolution using the widely adopted U-Net approach~\citep{ronneberger2015u, he2019learning}, applying it first to a theoretically well-understood prescription, the Zel'dovich approximation (ZA)~\citep{1970Zeldo}. Since the ZA is a simple linear dynamical mapping scheme, it has the benefit of providing a clear physical picture while enabling a comprehensive study of the neural network's physical interpretability. To further explore the physical benchmarking for more realistic nonlinear evolutions, we also test the same metrics on datasets generated by cosmological N-body simulations using the Particle-Mesh (PM) method, which has the benefit of being computationally inexpensive compared to higher-resolution approaches. Combining the ZA and PM methods, we are able to observe the behavior of the generative model on datasets with varying nonlinearity and derived from different algorithms to provide a more comprehensive benchmark of the neural networks' performance.

We pay significant attention to various metrics for judging the quality of the results from the generative model, focusing on those that have specific physical interpretations. It turns out that the choice of the loss function affects the results for these metrics in different ways and can have a very significant effect on the accuracy of certain outcomes, e.g., the mass fluctuation power spectrum, a key cosmological probe. While this is not unexpected, it does mean that the choice of the loss function is an important consideration when constructing the approximate generative maps.

The outline of the paper is as follows: We first introduce the physical formalism behind the generation of the training datasets, i.e., the ZA and PM methods for large-scale structure simulations (Section~\ref{sec:structure}), and then introduce the deep learning architecture and training method in Section~\ref{sec:ml}. This section also contains a detailed discussion of validation metrics and their physical implications. The training methodology is described in Section~\ref{sec:training}, where we begin with a conventional mean-squared error (MSE) loss and then describe and implement an improved density-weighted loss function. This section describes detailed notions of convergence and presents results for a number of performance metrics at the field level and for summary statistics; we also include a cross-power null test to verify the independence of the generative model results. Results for covariance matrices are presented and discussed in Section~\ref{sec:covariance}. We conclude by providing a summary of this work and discuss further implications in Section~\ref{sec:discussion}. 

\section{Cosmological Structure Formation}
\label{sec:structure}
In this section, we provide the background information for the structure formation study presented here, the evolution methods used, and how the training datasets are generated.
\subsection{Dynamical Evolution}
Dating back to the early universe, tiny perturbations in the matter density, possibly originating via quantum fluctuations from an epoch of cosmological inflation, constitute the seeds for the later formation and evolution of large-scale nonlinear structures. The gravitational Jeans instability amplifies the perturbations in an expanding universe, leading to the observed large-scale distribution of matter observed in galaxy surveys. 

The growth of structure can be treated via perturbative approaches~\citep{bernardeau2002large} when the density perturbations are small (i.e., the overdensity $\delta(\vec{x})\equiv (\rho(\vec{x})-\rho_b)/\rho_b$ is small compared to unity; here $\rho(\vec{x})$ is the local density and $\rho_b$ is the mean density of the Universe). Although perturbative techniques are limited in not being able to describe essentially nonlinear phenomena, they have the advantage of being analytically tractable and having well-defined dynamical properties. Thus, they are a useful test case for demonstrating certain strengths and weaknesses of DL-based generative models.

We use the ZA as a suitable perturbative approach because as a simple, yet powerful analytic technique, and as a Lagrangian method, it serves as a proper starting point towards building generative models based on N-body simulations. ZA-based particle evolutions are easy to generate and because the underlying trajectories are linear, they provide possibly the simplest target case for a neural network to capture.

\subsection{The Zel'dovich Approximation}
In Lagrangian perturbation theory, the key quantity of interest is the displacement field, which maps the initial particle position $\vec{q}$ into the final Eulerian positions $\vec{x}$ by %\citep{MartinWhite-ZA}
\begin{equation}
   \vec{x}(t) = \vec{q} + \psi(\vec{q},t), 
\end{equation}
% $$, $$
where $\psi(\vec{q},t_0) = 0 $. Every particle is uniquely labeled by its Lagrangian coordinate $\vec{q}$, and the displacement field $\psi(\vec{q})$ fully determines its motion. Lagrangian perturbation theory aims to expand the displacement field in terms of higher-order terms:
\begin{equation}
    \psi(\vec{q},t )= \psi^{(1)}(\vec{q},t)+ \psi^{(2)}(\vec{q},t)+ \psi^{(3)}(\vec{q},t) + ..., 
\end{equation}
% $$$$ 
where the ZA corresponds to the first-order solution, using the linear displacement field as the approximate solution for the dynamical equations. We note that the ZA is a local approximation -- the second-order correction adds in missing tidal effects.

Under the ZA, an initially uniform distribution given by Lagrangian coordinates $\vec{q}$ is displaced by:
\begin{equation}
    \begin{centering}
    \vec{x}(t)=\vec{q} + b(t) \vec{S}(\vec{q}),
    \end{centering}
\label{eq:co-moving}
\end{equation}
where $\vec{x}(t)$ are the comoving coordinates, $b(t)$ is the linear growth rate of fluctuations and $\vec{S}(\vec{q})= \vec{\nabla} \Phi(\vec{q})$ is the gradient of the initial gravitational potential $\Phi(\vec{q})$. The potential $\Phi(\vec{q})$ is determined by the primordial density fluctuations $\delta(\vec{x})$ via the Poisson equation, $\nabla^2 \Phi = 4 \pi G \delta$, where $G$ is the Newtonian constant of gravitation. Initial density perturbations are realizations of a Gaussian random field, which is fully specified by a (given) power spectrum. Particles representing mass elements (usually uniformly placed) are then moved according to the ZA. 

\subsection{N-Body Simulations: The Particle-Mesh method}
The PM method evolves the particle distribution by depositing particles on a spatial computational grid, thereby generating a density field, self-consistently solving the Poisson equation on the grid for the gravitational forces, and then stepping the particles forward in time using the self-consistent force given by the gradient of the gravitational potential that results from the solution of the Poisson equation. Symplectic time-stepping schemes are usually implemented using split-operator methods. PM methods are simple to implement, and their performance relies solely on the efficient solution of the Poisson equation. Typically, this relies on using Fast Fourier Transforms (FFTs), but other methods, such as multigrid, may be employed. In cosmological applications, PM codes are used when modest resolutions are sufficient to meet the intended purpose, as is the case here. 

We use \verb|FlowPM|~\citep{modi2021flowpm}, a GPU-accelerated PM N-body code built on Mesh-TensorFlow, for the generation of our training dataset.  \verb|FlowPM| is well suited for our purpose since it uses GPUs and can be implemented on the same system used to implement the DL-based generative model. \verb|FlowPM| is a distributed TensorFlow implementation for PM simulations; it uses a multi-grid implementation for force estimation based on multi-resolution pyramids and enables higher efficiency in PM data generation. 

\subsection{Generation of the Dataset}
 To test the ability of DL models to capture the linear and nonlinear evolution in cosmological simulations, we generated cosmological particle distributions using both the ZA and PM methods. The spatial dynamic range, while modest ($\sim100$) compared to N-body simulations (where it can reach $\sim10^6$), and to what might be required for most applications ($>10^3$), still improves significantly over the initial work of \cite{he2019learning} by roughly a factor of four.

In the case of the ZA, given an initial power spectrum $P(k)$, the displacement field $\textit{S}(\textit{q})$ is generated using an FFT-based technique, and the particles are moved from a regular lattice via ZA displacements. We generate 2000 pairs of ZA-evolved displacement fields at two different time steps of the evolution, with respective scale factors $a_0=0.0464$ (an ``early'' snapshot with redshift $z_0=20.5$) and $a_1=0.215$ (a ``late'' snapshot with redshift $z_1=3.6$). The evolution of the so-called ``cosmic web'' in between these two moments is given by the ZA (from Eq.~\ref{eq:co-moving}), solely determined by the initial potential gradient. %ZA is, in effect, ``stopped'' at the two subsequent quasi-linear stages with redshifts $z_0$ and $z_1$, and 
The displacement fields at these two times form the basis of the training/validation datasets. Data is split such that 1000 pairs are used for training, 500 for validation, and 500 for testing. Each realization is generated with a different random seed to ensure statistical independence.  %The dataset comprises of the displacement field of $64^3$ particles in a volume of (64 $h^{-1}$Mpc)$^3$.
\begin{figure*}[t]
    \centering
    \includegraphics[width=\textwidth]{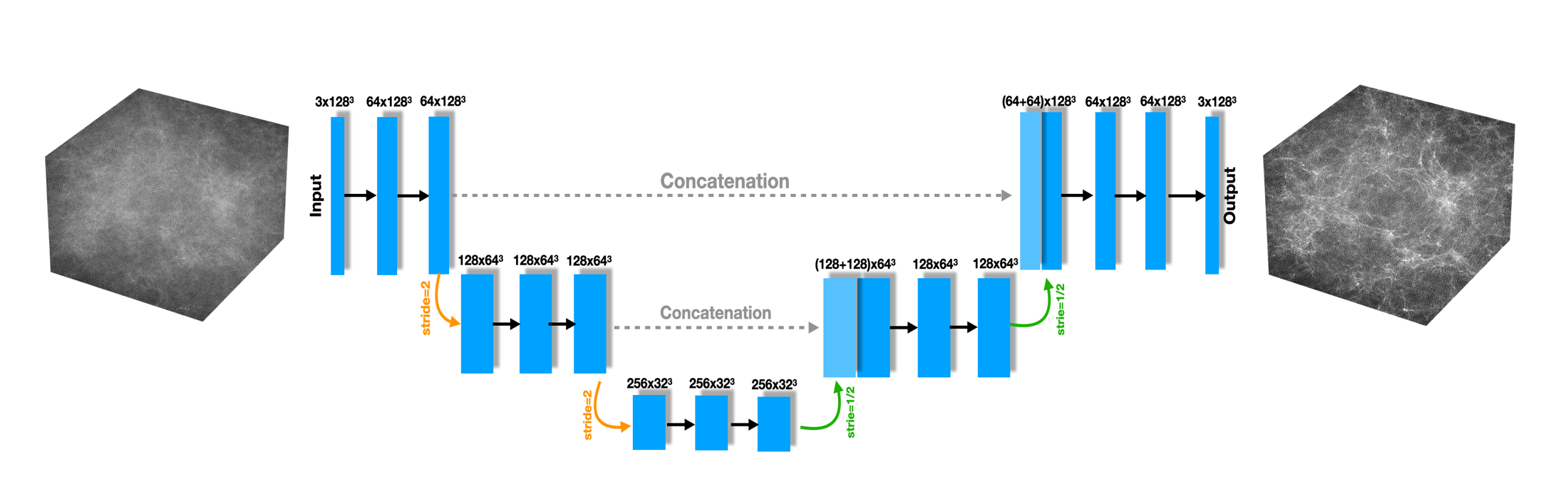}
    \caption{ Architecture for the 3-d U-Net showing the contracting and expanding parts. The input volume is progressively downsampled through multiple convolutional blocks (blue bars) with stride 2 (orange arrows), halving the spatial dimensions at each stage while increasing the number of feature channels. After reaching the bottleneck, the decoder path upsamples the feature maps via transpose convolutions (green arrows), concatenating them (dashed lines) with the corresponding feature maps from the encoder at each resolution level. This skip-connection strategy preserves high-resolution details lost during downsampling. The final output volume matches the original spatial dimensions of the input. }
    \label{fig:3d-unet}
\end{figure*}
The PM training set was generated by running simulations with a box size of 50 $h^{-1}$Mpc, evolving $128^3$ simulation particles using \verb|FlowPM|~\citep{modi2021flowpm}, on a GPU cluster. A total of 830 field realization pairs were generated, with 600 realizations generated for training and validation, and 230 designated for testing.   

In both cases (ZA and PM), for every simulation snapshot, we construct a three-channel volumetric field whose channels hold the Lagrangian displacement components of the $N^{3}$ simulation particles,
\[
  \mathbf{X}(\mathbf{q}) \;=\; \boldsymbol{\Psi}(\mathbf{q})
  \;=\; \bigl(\Psi_{x}(\mathbf{q}),\,\Psi_{y}(\mathbf{q}),\,\Psi_{z}(\mathbf{q})\bigr)
  \in \mathbb{R}^{3 \times N \times N \times N},
\]
where $\mathbf{q}$ denotes the Lagrangian grid index that permanently labels each particle.  
This tensor is fed directly into the U--Net {without any rescaling, normalization, or unit conversion}; the displacement amplitudes therefore remain in physical $\mathrm{Mpc}\,h^{-1}$ units.
Empirically we find that the number of training samples we selected allow effective capturing of the dynamical process from initial to final snapshots, which has also been confirmed by our convergence studies (details are provided in Section \ref{subsec:mse-conv}); preserving the absolute scale enables the network to learn an internally consistent forward map from the initial displacement field to the fully evolved field.

\section{AI-based snapshot translation framework and benchmarks}
\label{sec:ml}
A number of DL-based unsupervised generative models and supervised interpolation models have been applied in cosmological data creation~\citep{Ravanbakhsh2016,morningstar2018, Mustafa_2019, chardin2019deep, gunther2022cosmicnet}. Except for a few applications where the loss functions are tailored to the specific physical problem, most applications are domain-agnostic. That is, all the information about the underlying physics is entirely learned from the training data. These data-driven models have demonstrated reasonable accuracy in validation datasets, albeit with respect to metrics that closely resemble the loss function.

\subsection{Architecture}

Deep convolutional neural networks have been recognized for their exceptional performance in computer vision tasks, including pattern recognition, image classification, and segmentation. In this context, \cite{ronneberger2015u} proposed a convolutional neural network architecture, U-Net, that works well for biomedical image segmentation tasks, especially when training samples are limited.

The U-Net architecture, as shown in Fig.~\ref{fig:3d-unet}, is composed of a contracting path and an expansive path, both consisting of convolutional layers. The contracting path acts as a feature encoder and is made up of a sequence of 3-d convolutional layers. Each of these layers is followed by an activation layer (Rectified Linear Unit, or `ReLU') and stride 2 convolution downsampling layers. With each downsampling step, the number of feature maps doubles.

The expansive path serves as the decoder and is designed to mirror the contracting path. It incorporates upsampling layers to enlarge the spatial dimensions. This path is structured with upsampling layers followed by 3-d convolution layers. Each convolution reduces the number of feature maps by half and is concatenated with the corresponding feature maps from the contracting path. Subsequently, additional convolutional layers with activation functions are applied. In the upsampling segment, the substantial number of feature channels ensures the flow of context information to higher-resolution layers. Given the concatenations between feature maps, it is crucial to choose the input volume size judiciously, ensuring that the downsampling operations apply evenly across the $x$, $y$, and $z$ dimensions.

The use of U-Net was later extended from biomedical image segmentation to learning complex mappings between physical quantities during evolution~\citep{Giusarma2019, he2019learning, Calvo2019, Oliveira2020, Wu2021}. \cite{he2019learning} proposed a U-Net-based architecture to describe cosmic structure formation.  In order to make the U-Net capture the underlying physical symmetries more effectively, the padding and cropping procedures are modified to preserve the translation and rotational symmetries in the upsampling layers.

Here we adopt the 3-d U-Net architecture used in \cite{he2019learning} with 15 convolution and deconvolution layers. To determine optimal hyperparameters for the neural network layers, we conducted experiments on our dataset by varying the number of layers and latent dimensions in the U-Net model, ultimately selecting the architecture that yielded the best training performance, on which we report in this paper. %As stated earlier, the architecture has a contracting path and an expanding path. A basic block consists of a convolution layer, a batch normalization layer, and a ReLU layer. 

As can be seen from the architectural illustration (Fig.\ref{fig:3d-unet}), the basic composing unit for U-Net is transposed convolutional layers followed by ReLU activation and batch normalization layers. Starting from the input feature map, it goes through two 3x3 convolution layers with strides 1 and 2. The first few layers have an output number of channels from 64, 128 to 256, forming an expansive path. The set of five layers is each followed by a batch normalization layer and a ReLU layer.  They are then connected to a periodic padding layer, and the resulting middle layer gets chopped and concatenated with previous layers, then goes through a series of convolution layers with shrinking feature maps, leading to the final output result.

As an initial choice, we use the mean squared error (MSE) between the Lagrangian coordinates as the basis of the loss function, with L2 regularization, as 
\begin{equation}
\label{eq:loss}
    L = \frac{1}{N_p^3}\sum_{i=0}^{N_p} \sum_{j=1}^{3} (x^i_{j, \rm{true}}-x^i_{j, \rm{pred}})^2.
\end{equation}
Note that this simple form of the loss uses the common mean-squared-error formula and captures the distance from predicted Lagrangian coordinates versus the ground truth. For an actual physical system, this might not be the only loss function that we could implement; we will discuss other possible metrics to potentially use as the learning loss function later -- metrics that can help capture more detailed information.

\subsection{Validation metrics}
\label{sec:validation}
Beyond qualitative inspection, we now turn to listing a set of measures that quantify the nature of the matter distribution in the universe and the topological connectivity of large-scale structures. We also include one metric introduced specifically to look for artificial correlations (potentially) induced by the training protocol. These measures form the basis for metrics that will be used to assess the performance of the AI-generated forward map.

\subsubsection{Pixel-wise Comparison: The Density PDF}
\label{sec:pixel}
To cross-check the generated particle field from the neural network with ground truth simulation results, the most straightforward comparison is the relative error in predicted densities or particle displacements. Since the data we use for training and generation are the displacement fields of particles, to convert the displacement field of different particles (i.e., their $x$, $y$, and $z$ displacements) to a density field, we use the Cloud-In-Cell (CIC) deposition method to generate a density field on a regular grid. With a density field $\rho(\textbf{x})$ in hand, we define a local relative error field via
\begin{equation}
\delta\rho(\textbf{x}) = \frac{| \rho(\textbf{x})_{\rm{pred}} - \rho(\textbf{x})_{\rm{true}}|}{\rho(\textbf{x})_{\rm{true}}},
\end{equation}
where $\rho(\textbf{x})_{\rm{pred}}$ is the U-Net prediction, and $\rho(\textbf{x})_{\rm{true}}$ is the ZA or PM result.  

In addition to the field-level information, we also use the one-point probability distribution function (PDF) of the density field to assess prediction fidelity via a convenient summary statistic. Although this PDF is not sensitive to clustering properties of the field, it is sensitive to how well the dynamic range in density -- an important quantity -- is being reproduced.

\subsubsection{Matter Power Spectrum}
The power spectrum of density fluctuations is a statistic of central significance in cosmology as it robustly describes the clustering of matter in the universe~\citep{Peebles1980, peacock1996non}. The power spectrum is the Fourier transform of the two-point correlation function in real space. Denoting the matter overdensity as $\delta (\textbf{x})= ({\rho(\textbf{x})-\overline{\rho}})/{\overline{\rho}}$, where $\overline{\rho}$ is the mean density, and writing its Fourier transform dual as $\delta(\textbf{k})$, the power spectrum $P(k)$ is defined by %Equation~\eqref{eq:pk} 
\begin{equation}
\label{eq:pk}
\langle \delta(\textbf{k}) \delta(\textbf{k}^{\prime}) \rangle= (2\pi)^3 P(k) \delta^3(\textbf{k}-\textbf{k}^{\prime}).
\end{equation}  
The evolution of structure in the universe is driven by the Jeans instability under which initially all modes grow independently, until eventually nonlinear effects become important. In the power spectrum, this is reflected in a uniform growth over time, with nonlinear effects entering at a wavenumber $k_{NL}$ that moves from higher values to lower as the redshift decreases (or as the scale factor increases).

\subsubsection{Higher order Correlation -- Bispectrum}

The power spectrum is a measure of two-point statistics; by itself, it is not a sufficient probe of non-Gaussianity induced by evolution under gravity. Natural extensions involve higher-point statistics such as 3-point (or higher) and are especially useful for studying the late stages of structure formation where the evolution is highly nonlinear and non-Gaussian. 

The bispectrum, the Fourier equivalent of the three-point correlation function, is a good metric for benchmarking. In recent years, significant research efforts have been devoted to bispectrum studies, especially for small departures from Gaussianity in the primordial cosmological perturbations~\citep{sefusatti2010matter}. Higher order statistics can break the degeneracies between bias and cosmological parameters, lift the degeneracies for primordial non-Gaussianities, and the combined studies of bispectrum with power spectrum helps reveal more cosmological large-scale structure information~\citep{hashimoto2017precision}, and tighten constraints on dark energy and modified gravity through redshift-space distortions. 

Adapting similar symbol conventions as above, the bispectrum  $B(k_1,k_2,k_3)$ \citep{hung2019advancing} is defined as 
\begin{equation}
\label{eq:bis}
\langle \delta(\textbf{k}_1) \delta(\textbf{k}_2)\delta(\textbf{k}_3) \rangle= (2\pi)^3 B(k_1,k_2,k_3) \delta^3(\textbf{k}_1+\textbf{k}_2+\textbf{k}_3). 
\end{equation}
Given the limited purposes of the current work, we will focus attention on the equilateral triangle case for simplicity.

\subsubsection{Topological metrics -- Percolation Analysis}
\label{sec:perc}
Two-point functions and power spectra are robust clustering measurements in both observations and simulations, but they do not contain shape or topological information. While a full ensemble of $n$-point correlation functions or their Fourier space equivalents contains, in principle, complete information of the spatial distribution of cosmic structures, this information is nontrivially distributed across the $n$-point functions. These metrics can also be computationally expensive to compute for a large number of samples, such as simulation realizations. Hence, several higher-order indicators specific to particular physical, morphological, or connectivity phenomena have been proposed to characterize the structure of the cosmic web. These include minimal spanning trees \citep{Barrow1985}, genus curves \citep{Gott1986},  excursion set approaches for modeling voids \citep{Shandarin2006}, Minkowski functionals for characterizing the shapes of individual regions \citep{Sahni1998}, excursion sets of density fields \citep{Percolation} and Morse-Smale complexes in the density fields \citep{Sousbie2011, Shivshankar2015}. 

Since connectedness of dark matter density clusters is associated with the emergence of the cosmic web, and percolation properties can be described by simple scaling relations, we choose percolation as a topological metric that quantifies the connectivity of spatial structure within the cosmic web (\citealt{Percolation}). This statistic models fragmentation, connectivity, and persistence of percolation. First, we calculate the volume of the excursion set $V_{ES}(\rho/\bar\rho)$, the region with a lower bound on overdensity.
The volume fraction of the excursion set $f_{ES}(\rho/\bar\rho) =  {V_{ES}(\rho/\bar\rho)}/{V_{tot}}$ with respect to total volume is computed across varying thresholds of overdensity values by identifying the regions with densities exceeding the threshold. The volume fraction of the largest structure  $f_{1}(\rho/\bar\rho)$ is also computed for the same overdensity threshold. The filling fraction $f_1/f_{ES}$ as a function of $f_{ES}$ contains information about the topological phase transition.

\begin{figure}
  \centering
  \includegraphics[width=0.5\textwidth]{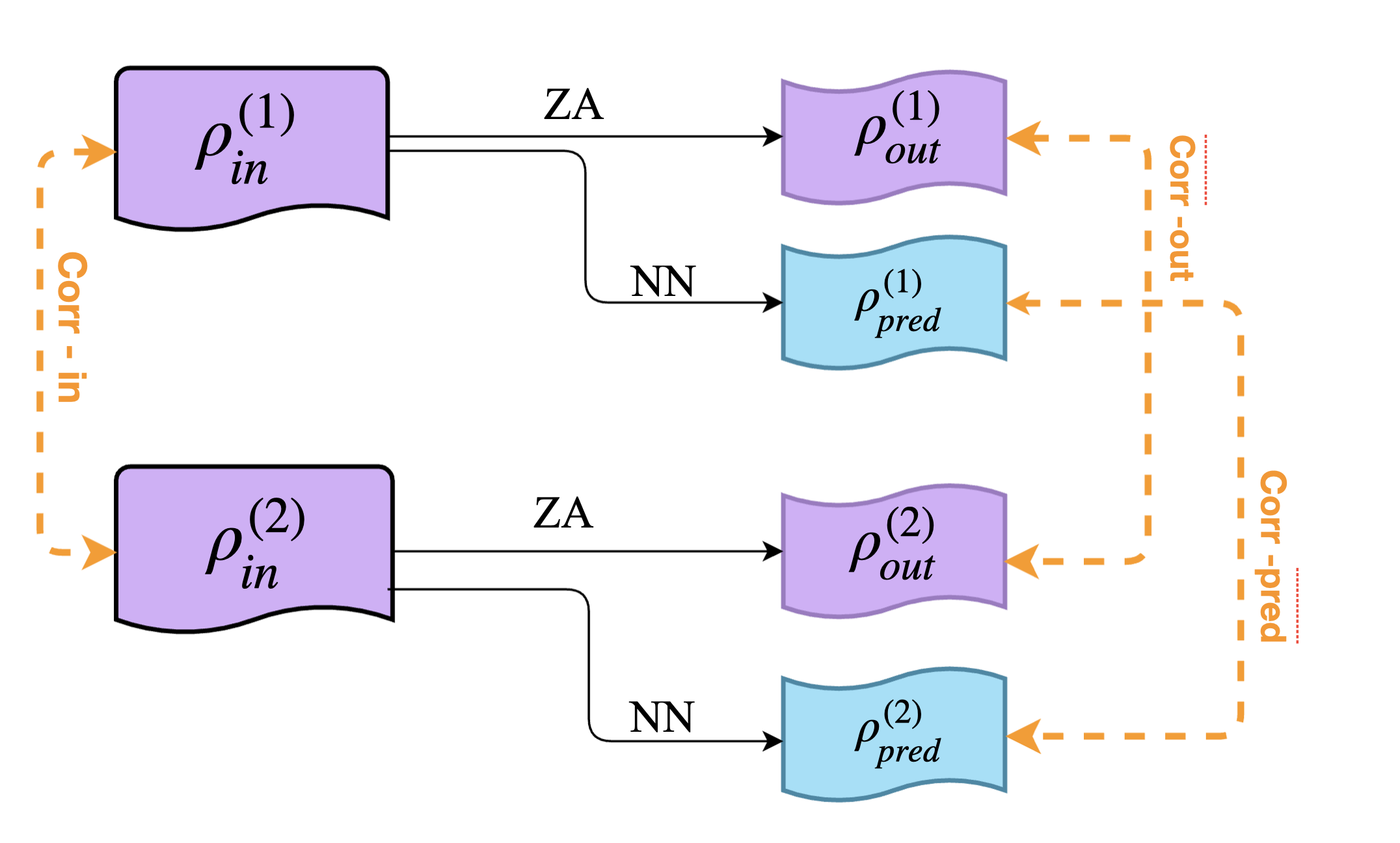}
  \caption{Illustration of the cross-power test (Section~\ref{cross}), showing the density fields and cross-correlation power spectra computed between them.}
  \label{fig:crosspow-illustration}
\end{figure}

In the matter density field, the excursion set normally consists of a number of isolated fragments with different volumes. The volume fraction $f_{ES}(\rho/\bar\rho)$ increases with reducing the overdensity limit. The largest isolated region and the corresponding volume fraction $f_1$ are easily computed in a numerical simulation via voxel counting in the density field. When the filling fraction $f_1/f_{ES}$ is close to one, the largest isolated structure occupies most of the excursion set. This signifies the existence of a single percolating structure through most of the cosmic field. When the filling fraction is close to zero, none of the isolated structures dominate the excursion set. This represents fragmented structures in the cosmic field. In the case of the matter density field, the filling fraction $f_1/f_{ES}$ grows from zero to unity with decreasing overdensity values ($f_{ES}$ functions as a proxy for the density threshold). In the case of $\Lambda$CDM, as $f_{ES}$ is increased, a percolation transition occurs in a smooth, but relatively sharp manner and at significantly lower $f_{ES}$ values as the field evolves, becoming more nonlinear -- a feature present in the ZA as well as in full N-body runs~\citep{Percolation}.

\subsubsection{Cross-Power Test}
\label{cross}
Cosmic evolution as a physical process follows fixed dynamical rules, independent of the initial conditions. However, the neural network emulating the evolution might show discernible bias inherited from the finite sampling over a limited set of examples -- an inherent property of a finite training dataset. Specifically, we investigate whether the U-net prediction induces otherwise non-existent correlations among the outputs of independent realizations, in contrast to the ZA or PM-evolution of fields, where independent initial conditions result in independently evolved fields.

To carry out this test, we use the cross-power spectrum across two different density fields following the scheme outlined in Fig.~\ref{fig:crosspow-illustration}.  We first generate two initial conditions, independent of the training set, measure their cross-power spectra, then evolve them separately by ZA/PM and U-Net, and then measure the output cross-power between them again. By comparing the final cross-power spectra we can see if there are any generative model-induced correlations in the NN results as compared to the final cross-power given by ZA or PM, both of which result from two independent evolution maps acting on the initial conditions.

%%%%%%%%%%%%%%%%%%%%%%%%%%

\section{Model Training}
\label{sec:training}
During training, as previously described, the model is fed with randomly selected pairs of initial and final displacement fields -- derived from both ZA and PM evolutions -- to learn the mapping from the initial field (at redshift, $z=z_0$) to the final field configuration (at redshift, $z=z_1$). After training, the model is evaluated on independent test datasets (separate from the training and validation sets) to compare its predictions against the ground truth.

Throughout the training, we used the iterative Adam optimizer of~\cite{kingma}, with learning rate 1e-5, $\beta_{1,2}=(0.9, 0.999)$, and weight decay regularization 1e-5. The model was saved every 500 steps and evaluated by the validation dataset every 20 steps (more details below). We used the validation dataset to optimize the hyperparameter choices for best training results, and also compared with commonly used values from similar models in the literature. Model training was carried out on the Argonne Laboratory Computing Resource Center (LCRC) Swing cluster -- a single node of Swing has 8 NVIDIA A100 GPUs with a combined memory of 320GB. The training of each model for a specific redshift pair (for either ZA or PM) takes around 66 hours of total wall clock time. 

\label{gen_inst}
\begin{figure}
    \centering
    \includegraphics[width=0.52\textwidth]{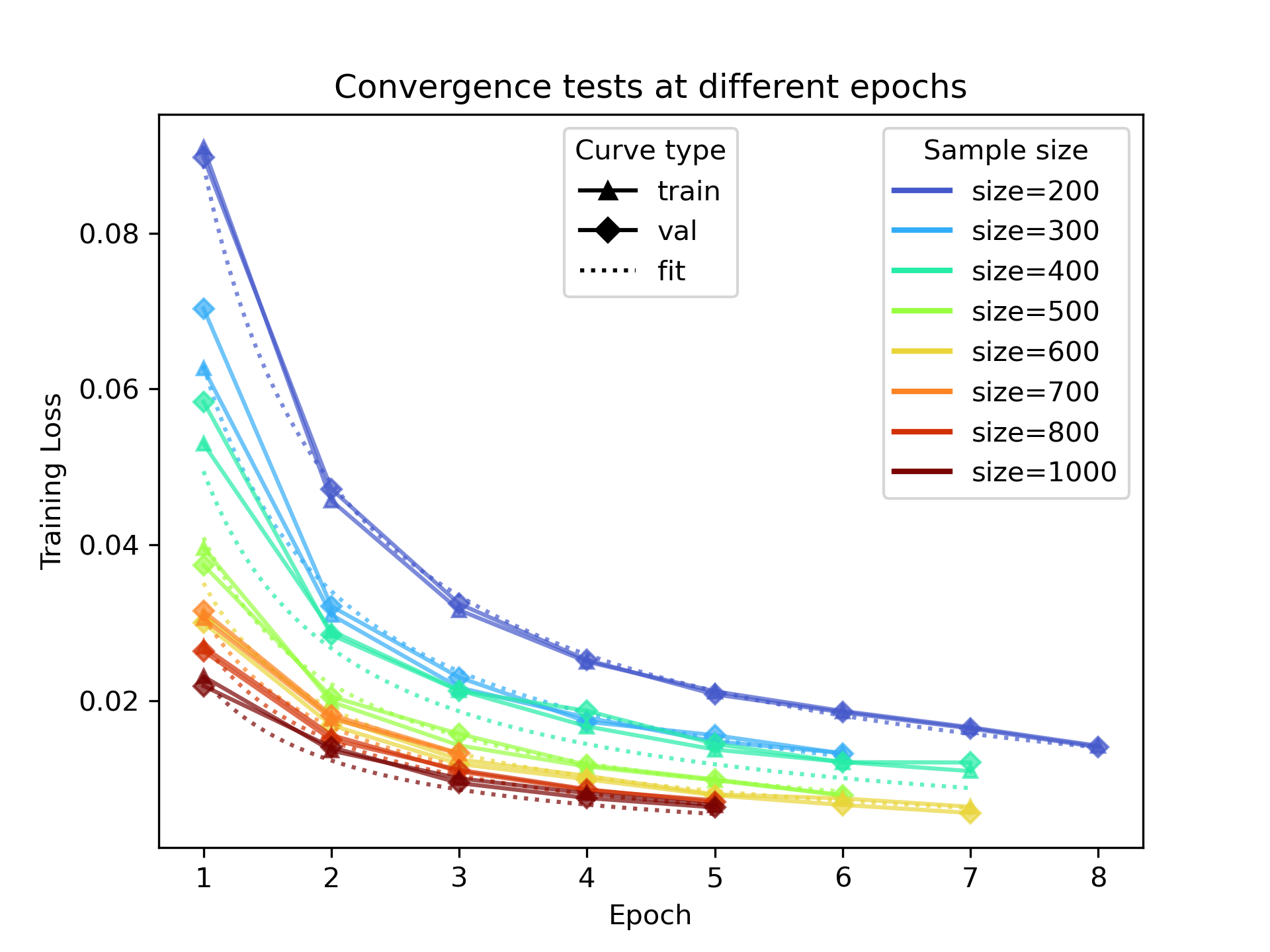}
    
    \includegraphics[width=0.52\textwidth]{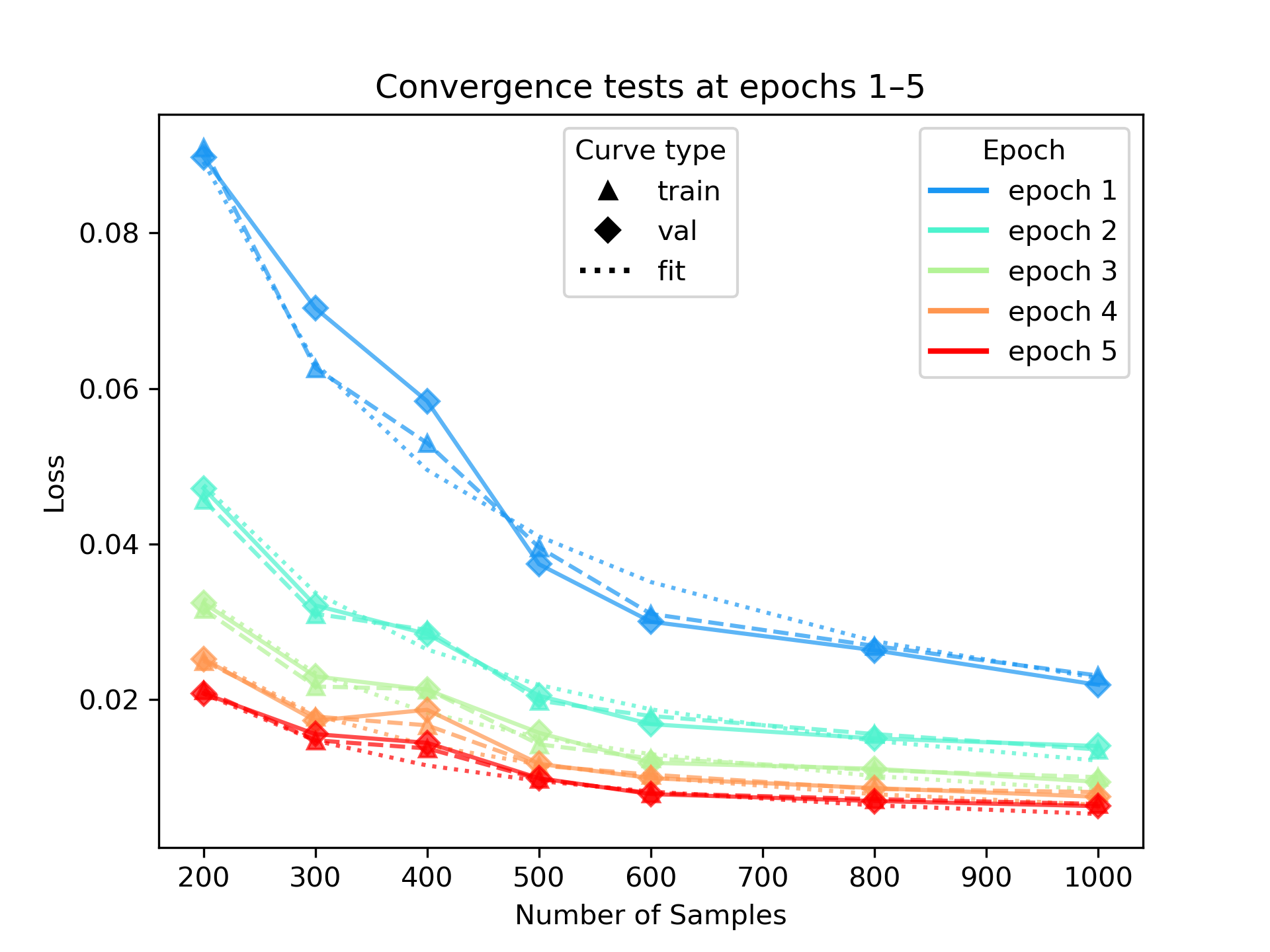} 
    \caption{Convergence investigations with sample size and epoch (Section~\ref{subsec:mse-conv}). Top panel: Training and validation loss curves as a function of epoch number, for different-sized training datasets. Bottom panel: The first five epochs during the training, validation and training losses are plotted as a function of the number of samples in the training set.} 
    \label{fig:convergence}
\end{figure}

\begin{figure*}
     \centering
         \includegraphics[width=\textwidth]{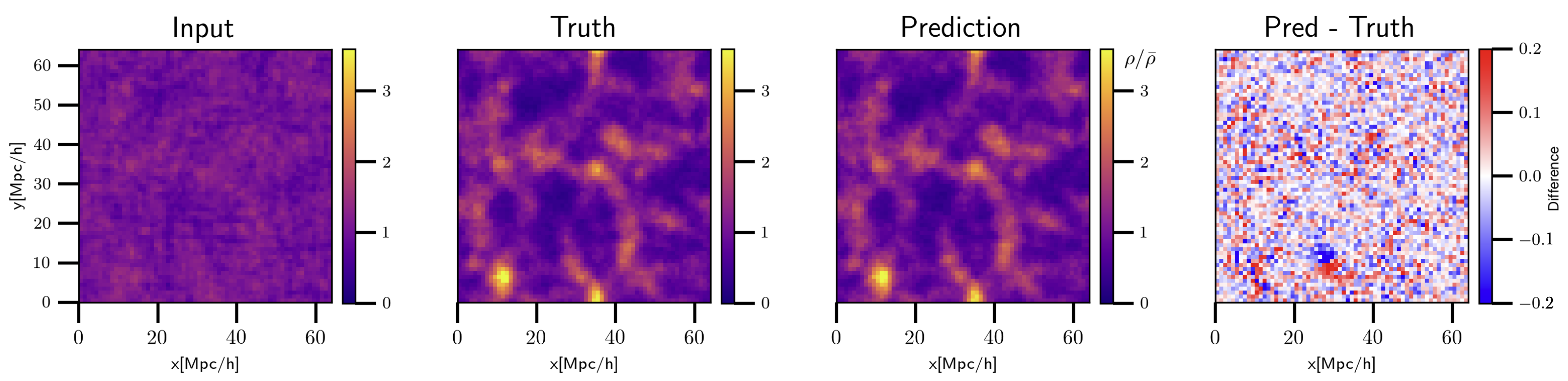}
         \caption{Comparison of projected densities between the predicted cosmic web and ZA ground truth. Densities are derived from the displacement field with a Cloud-In-Cell (CIC) method and summed over one axis. The `input' panel (first panel) presents the density field projection of the early ($z=z_0$) snapshot. The color bar in each panel shows the magnitude of the matter density, $\rho(\textbf{x})$. In the above case, the two scale factors are $a_0=0.0465$ and $a_1=0.215$, and the box size is 64 $h^{-1}$Mpc, with $64^3$ particles.}
         \label{fig:dens-ZA}
\end{figure*}

\begin{figure*}
     \centering
         \includegraphics[width=\textwidth]{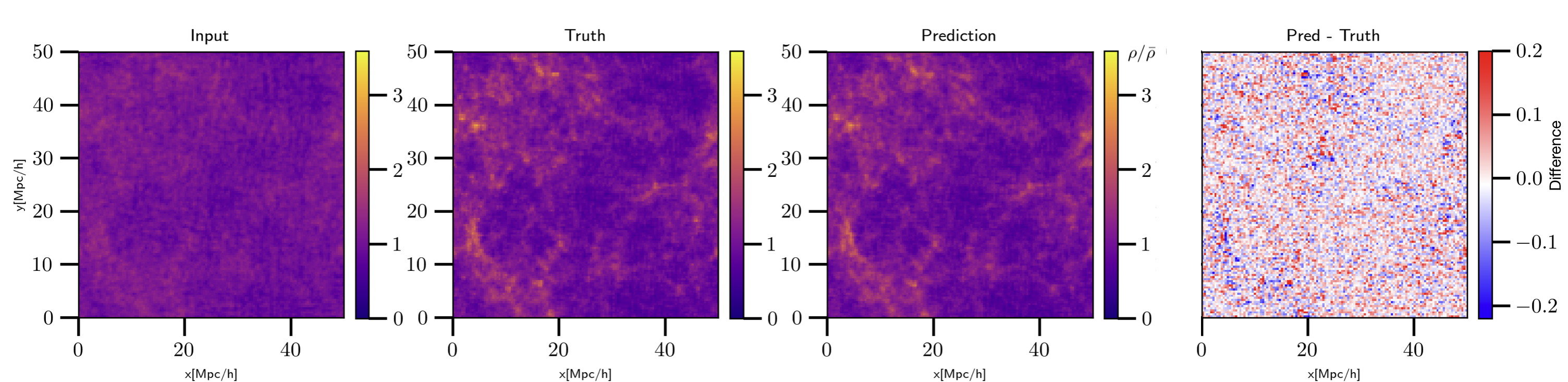}
         \caption{Comparison of projected densities between the predicted cosmic web and PM-evolved ground truth. $\sim 600$ samples are used for the training, each of them containing $128^3$ particles in 50~$h^{-1}$Mpc boxes. Densities are derived from the displacement field with a CIC method and summed over one axis. The `input' panel (first panel) presents the density field projection of the early ($z=z_0$) snapshot. The color bar in each panel shows the magnitude of the matter density $\rho(\textbf{x}$). The snapshots used are from $a_0=0.05$ and $a_1=0.1$.}
         \label{fig:densPM} 
\end{figure*}

\subsection{Results with native loss function training}
We carried out the training and evaluation protocol in two steps, first with the widely-used conventional mean-squared error (MSE) loss of Eq.~\eqref{eq:loss}. Motivated by the initial results, we followed up by repeating the training using a density weighted loss function (described later below), which yielded significantly improved results, especially at smaller length scales.

Our \emph{baseline} model is trained with the MSE objective of Eq.~\eqref{eq:loss}; i.e., each Cartesian
component of the predicted displacement field is compared \emph{one-to-one}
with the ground-truth field, and the resulting squared differences are
averaged over \emph{all} particles and dimensions without any additional
spatial or scale-dependent weighting.  Consequently, every voxel contributes equally to the global loss and the optimiser is driven to reproduce the
\emph{volume-averaged} behaviour of the field.

In practice we trained for $\sim$ $10,000$ optimiser steps
($\sim 50$ epochs for the ZA data and $\sim 35$ epochs for the PM data) with a batch size of~8 using the \textsc{Adam} optimiser
(\texttt{lr}=$10^{-5}$, $\beta_1=0.9$, $\beta_2=0.999$) and an
$L_2$~weight-decay of~$10^{-5}$.  A cosine-annealing learning-rate
scheduler with a 500-step warm-up stabilises the early stages of training.
Checkpoint models are written every 500 steps and the validation loss is
monitored every 20 steps to guard against over-fitting.

This simple MSE formulation follows the precedent set by most image-to-image translation studies in cosmology (e.g., \citealt{Ravanbakhsh2016,Mustafa_2019,he2019learning,Giusarma2019,Wu2021}), and therefore provides a useful reference point for assessing the impact of the density-weighted loss introduced in the next subsection.

\subsubsection{MSE loss convergence with sample size and epoch}
\label{subsec:mse-conv}
Deep learning predictions are tied to the information contained in training datasets, and the size of the training dataset thus has a significant impact on the results obtained.  If the neural network can capture the underlying physical dynamics of the evolution sufficiently well, then, at some point the size of the training dataset used should cease to matter; up to this point we expect to see (some notion of) improved convergence as the training set size increases. It is important to understand the size of the training ensemble at which point an acceptable accuracy for the target metrics is achieved. If convergence is not achieved early enough with sample size, the training protocol may become computationally too expensive for the problem at hand.

To understand how the size of the training set influences the effectiveness of training, we conduct a convergence test of different training sizes $N_{\rm train}$, ranging from 100 to 1000 samples. The testing and validation set for these training schemes are kept the same, different from the varying training datasets. We start with considering the behavior of the MSE loss.

The MSE losses for training and validation datasets versus training size at different epochs are plotted in Fig.~\ref{fig:convergence}. In the top panel, we show the training loss curves of different sample sizes as a function of number of epochs;  if we focus on a certain epoch and plot the training/validation losses over training set sizes, we can see the effect of sample number on the loss curve, which is shown in the bottom panel.

By fitting the loss in terms of training data-size $N_{\rm train}$ and training epoch $E_{\rm train}$, we acquire an approximate power-law scaling formula as follows (see Fig.~\ref{fig:convergence}):
\begin{equation}
{\rm loss} = 7.9 {E^{-0.90}_{\rm train}} {N^{-0.84}_{\rm train}}.
\label{eq:loss-scaling}
\end{equation}

According to this convergence fitting formula, for a fixed number of training epochs, the loss scales with the number of training samples as a power law with an index of approximately $-0.84$. The extent of the convergence tests was limited by computational queue time restrictions -- especially for a larger number of samples, it takes approximately linear ($\mathcal{O}$(N)) more training time to finish the same number of epochs, and thus leading to a limited number of epochs for our convergence plots. While we aim to improve this situation in future, the general conclusions arrived at here are sufficient for the purposes of this paper.

We now present some initial qualitative results for the U-net predicted fields. Once the loss has sufficiently converged (after around 10 training epochs), we first consider the quality of the reconstructed cosmic web. Figure~\ref{fig:dens-ZA} shows the generated density (projected along the $x$-axis) of the cosmic density field for one test initial condition, demonstrating a reasonably good agreement with the ZA result. A similar demonstration for the PM runs is shown in Fig.~\ref{fig:densPM} (again with around 10 epochs of training). For both PM and ZA visual density comparisons, the reconstruction of density fields is qualitatively in good agreement with the reference; the two, when subtracted, generate an approximately random field although some `hot spots' are visible in areas of higher density contrast. More detailed quantitative benchmarks will be presented in Section~\ref{sec:densweight}.

\subsubsection{Convergence with snapshot redshifts}

Aside from the training data size and training epoch, the effective dynamic range also needs to be considered. As structure evolves from an earlier to a later stage, the density contrast increases and nonlinearity in the spatial clustering of matter is enhanced. Thus the mapping between the initial snapshot and the final displacement field becomes more complex as the final time is increased (or as the final redshift is made smaller). 

To test the deep neural network's capability to capture this increasingly complex mapping, we conduct a convergence study for different final snapshot redshifts.
For both the PM and ZA runs, we generate the datasets at a sequence of redshifts, marked by their different scale factors $a = 0.1, 0.2, 0.4, 0.6, 0.8 $ and $1.0$. We then train the models to capture the mapping between different pairs from initial and later snapshots. To test the capability of the model as simply as possible, we keep training setups, hyperparameters, and architecture the same for these choices. The only differences are the choices of training dataset available at different redshifts.

\begin{figure}
    \centering
    \begin{overpic}[clip,width=0.5\textwidth]{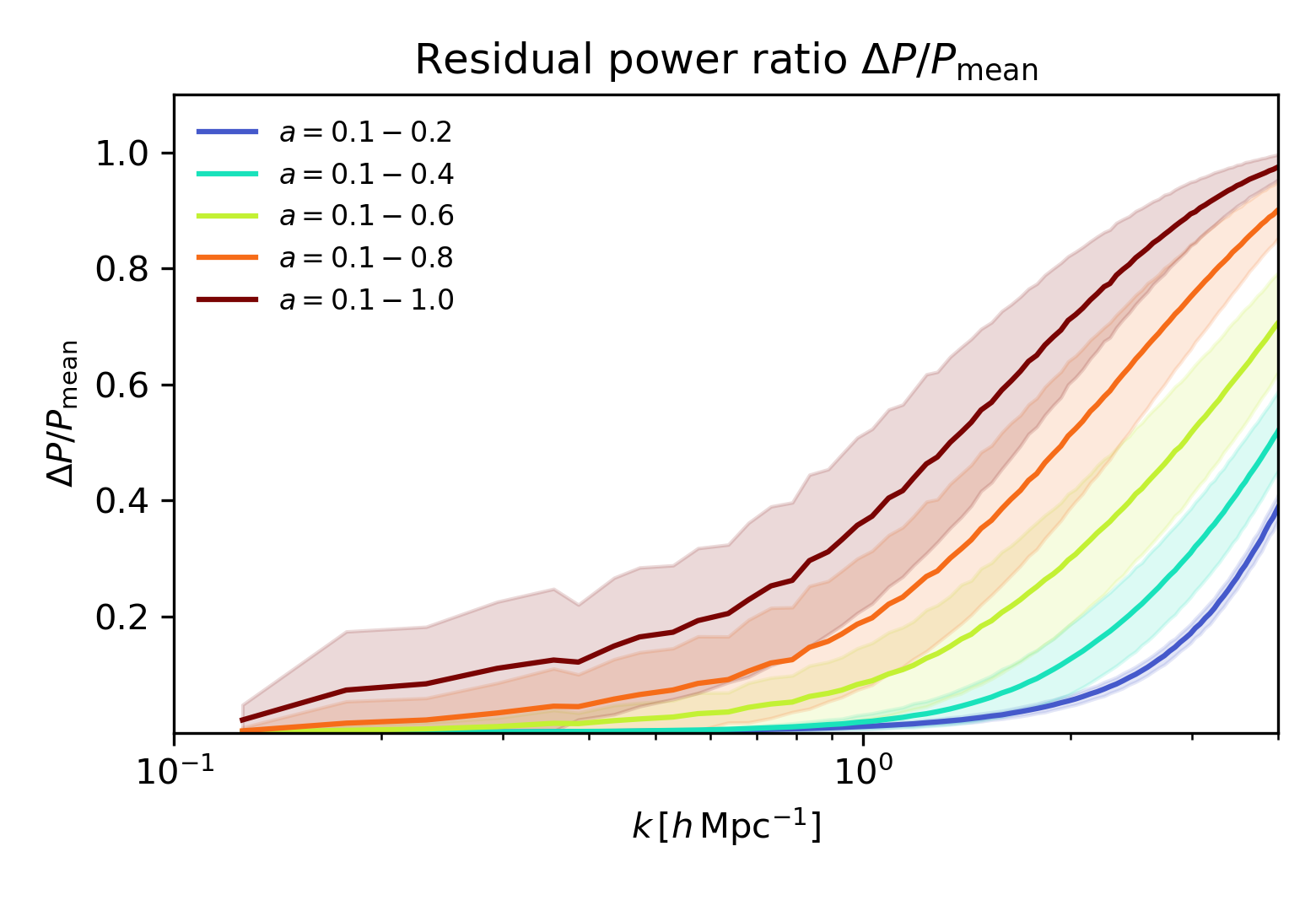} 

    \end{overpic}
    
\vspace{-.3cm}

    \begin{overpic}[clip, width=0.5\textwidth]{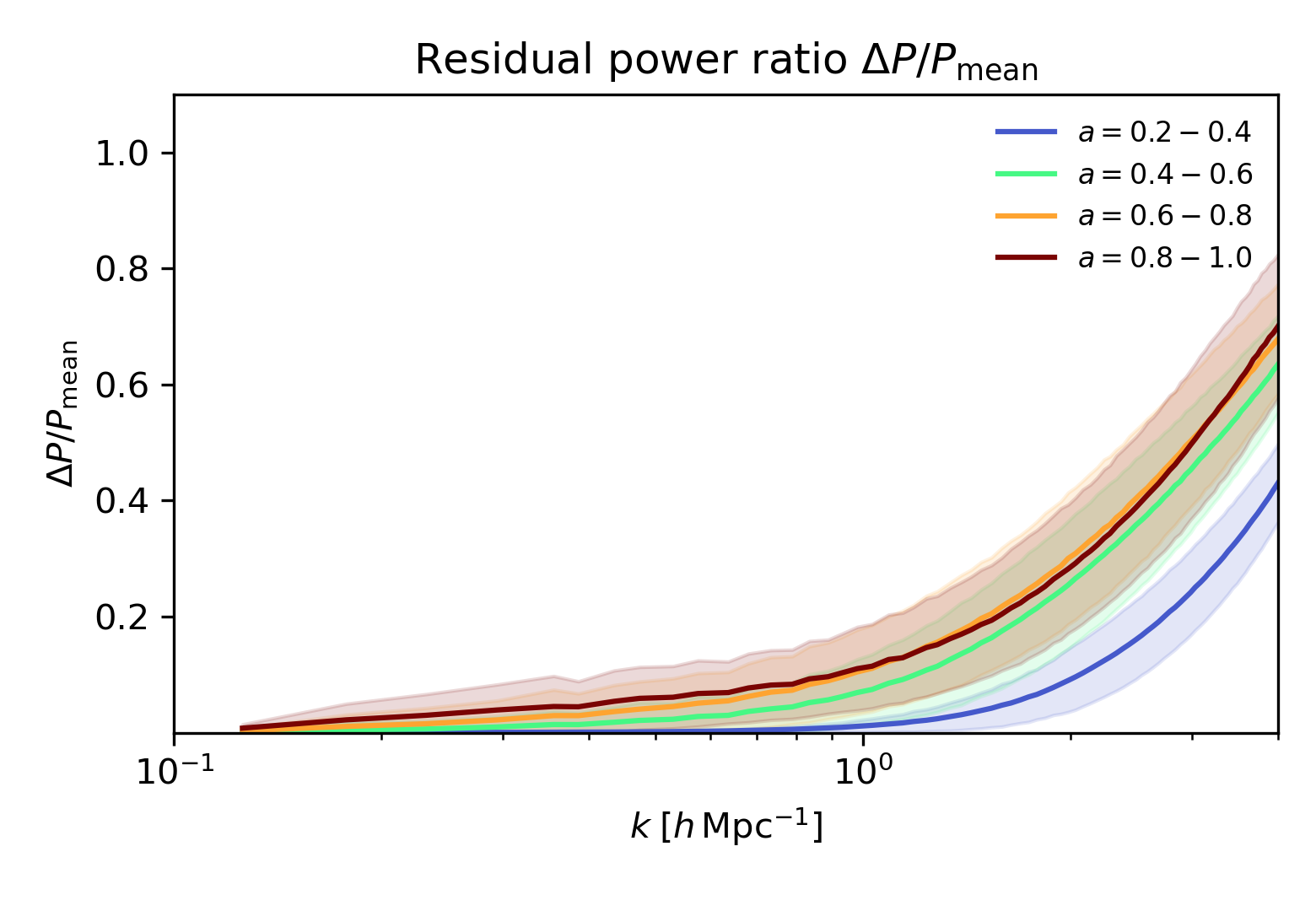}
        
    \end{overpic}
    
\vspace{-.5cm}

\caption{Ratio of residual to true power spectra, $\Delta P / \bar{P}$, for density fields predicted by the U-Net emulator. 
{Top panel:} Initial snapshot fixed at $a = 0.1$; curves show final targets $a = 0.2,\;0.4,\;0.6,\;0.8,\;1.0$. 
{Bottom panel:} Redshift interval fixed at $\Delta a = 0.2$; curves correspond to start–end pairs $(0.2 \rightarrow 0.4),\;(0.4 \rightarrow 0.6),\;(0.6 \rightarrow 0.8),\;(0.8 \rightarrow 1.0)$. 
In both panels the residual ratio increases with larger look‐back intervals and with later starting epochs, indicating that prediction errors grow with both the redshift gap and the cosmic time at which the evolution begins.}
\label{fig:z_dependence_power}
\end{figure}

For making the effect of dynamic range more apparent, we compare the power spectrum of the {\em residual} field as a function of different $\Delta a$ values (Fig. \ref{fig:z_dependence_power}). we can observe an increase with respect to redshift gap change, and initial snapshot redshift change. In the plot, we show the power spectrum computed from the absolute value of the residual field (prediction-ground truth). At the largest redshift gap ($a=0.1$ to $1.0$), the residual power spectrum of the absolute value of the over-density field increased by two orders of magnitude compared to the smallest redshift gap ($a=0.1$ to $0.2$). The implication of this result is that generative models -- depending on the accuracy required -- will likely need a number of intermediate training results in order to maintain control on accuracy metrics across the full required range of scale factor (or redshift).

\subsubsection{Convergence for error displacement fields}
\label{subsec:convg_for_displacement}

During training, the neural network iteratively adjusts its parameters to minimize discrepancies between its predicted particle displacement distributions and the ground truth.  Initially, the parameters are randomly initialized, but through gradient descent and back-propagation of the mean-squared error (MSE) loss computed between the predicted and true displacements, they gradually converge toward the correct mapping. As shown in Fig.~\ref{fig:convergence}, the average discrepancy over all particle displacements decreases smoothly with increasing training epochs. However, a reduction in the overall loss does not guarantee that every individual prediction converges monotonically toward its ground truth; some individual displacement predictions may even deteriorate as training progresses.

To gain deeper insight, we examine the statistical distribution of individual errors throughout the training process. For each particle, at each training step, the ground truth displacement vector and predicted displacement define two distinct fields. The instantaneous error can be quantified by the 2-norm of the difference between these two vector fields: $(|\mathbf{A} - \mathbf{A}'|_{2-norm})$.  Here, \(\mathbf{A} = (x, y, z)\) represents the ground truth displacement, and \(\mathbf{A}' = (x', y', z')\) denotes the NN-predicted displacement. This 2-norm denotes a spatially varying ``error field''. The displacement error is computed as:
\begin{equation}
|\mathbf{A} - \mathbf{A}'| = \sqrt{(x - x')^2 + (y - y')^2 + (z - z')^2}.
\label{err_conv}
\end{equation}

To evaluate the error convergence behavior of the NN predictions during the process of training, we analyze the statistical distribution of displacement error \(|\mathbf{A} - \mathbf{A}'|\) throughout the training process. For the statistics of the error field, we naturally select a few representative statistics, namely the maximum, average, and group-sampled displacement errors from the error field. The dataset we selected is from the $a=0.1$ to $a=0.2$ snapshot pairs.

\begin{figure}
    \centering
    \includegraphics[width=0.5\textwidth]
    {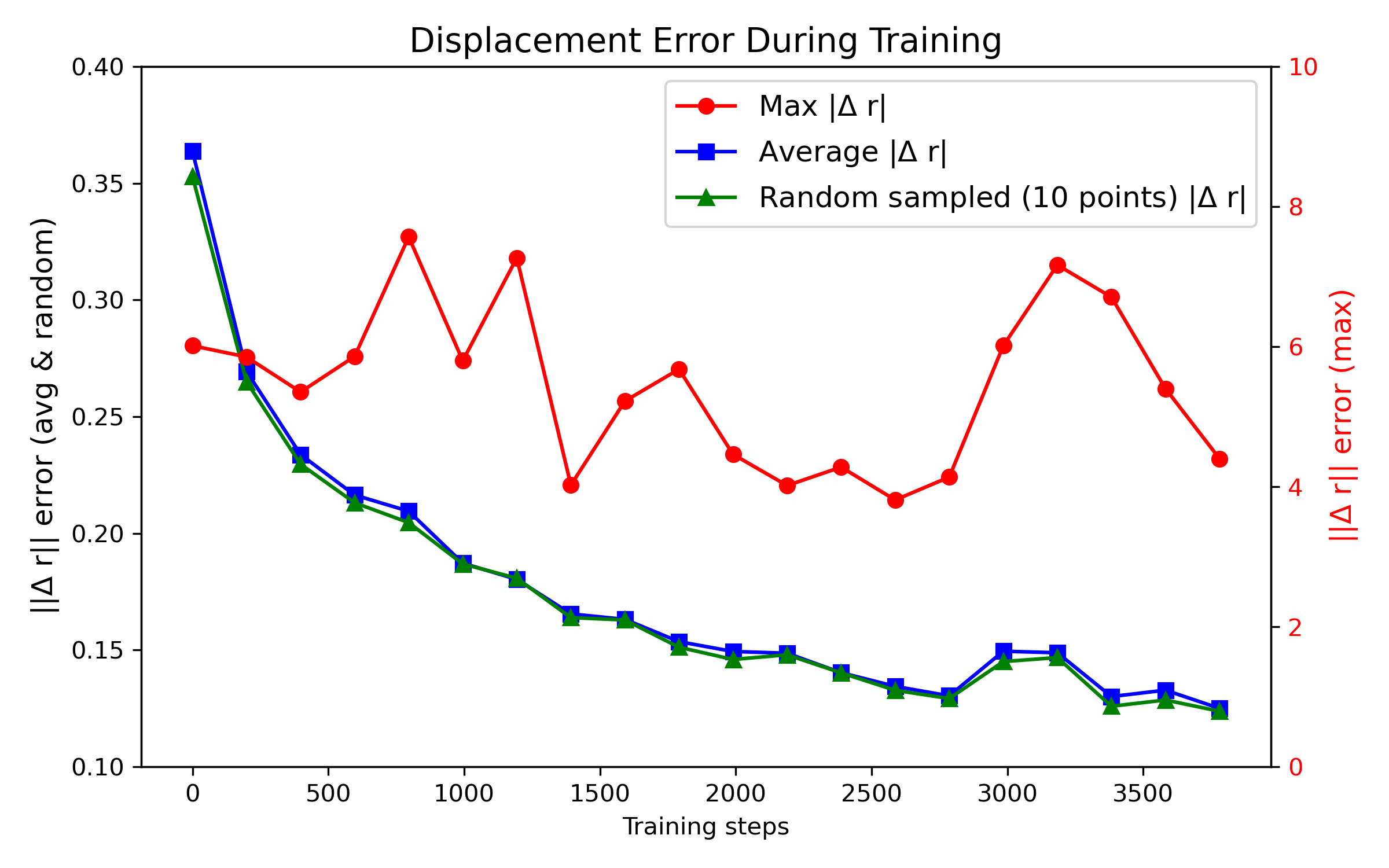}
     % \label{fig:displacement_metrics}
    \caption{Error convergence behavior of the displacement error field during training for the unweighted MSE. The plot shows the evolution of a few statistics of the displacement error field (Eq.~\ref{err_conv}) during the training process: the maximum, average, and mean of from 10 randomly selected points. The training dataset is for the $a=0.1$ to $a=0.2$ snapshot pair.}
    \label{fig:displacement_metrics}
\end{figure}

In Fig.~\ref{fig:displacement_metrics} we plot the maximum, average, and randomly sampled (10 points) displacement errors over multiple validation checks. These metrics provide insights into the convergence behavior of the displacement field outputted by the NN model during training. From the analysis we observe that the average displacement MSE error converges,  while the maximum among the distribution of displacement errors does not appear to uniformly converge to zero, but rather exhibits large fluctuations. 

\begin{figure}
    \centering
    \includegraphics[width=\linewidth]{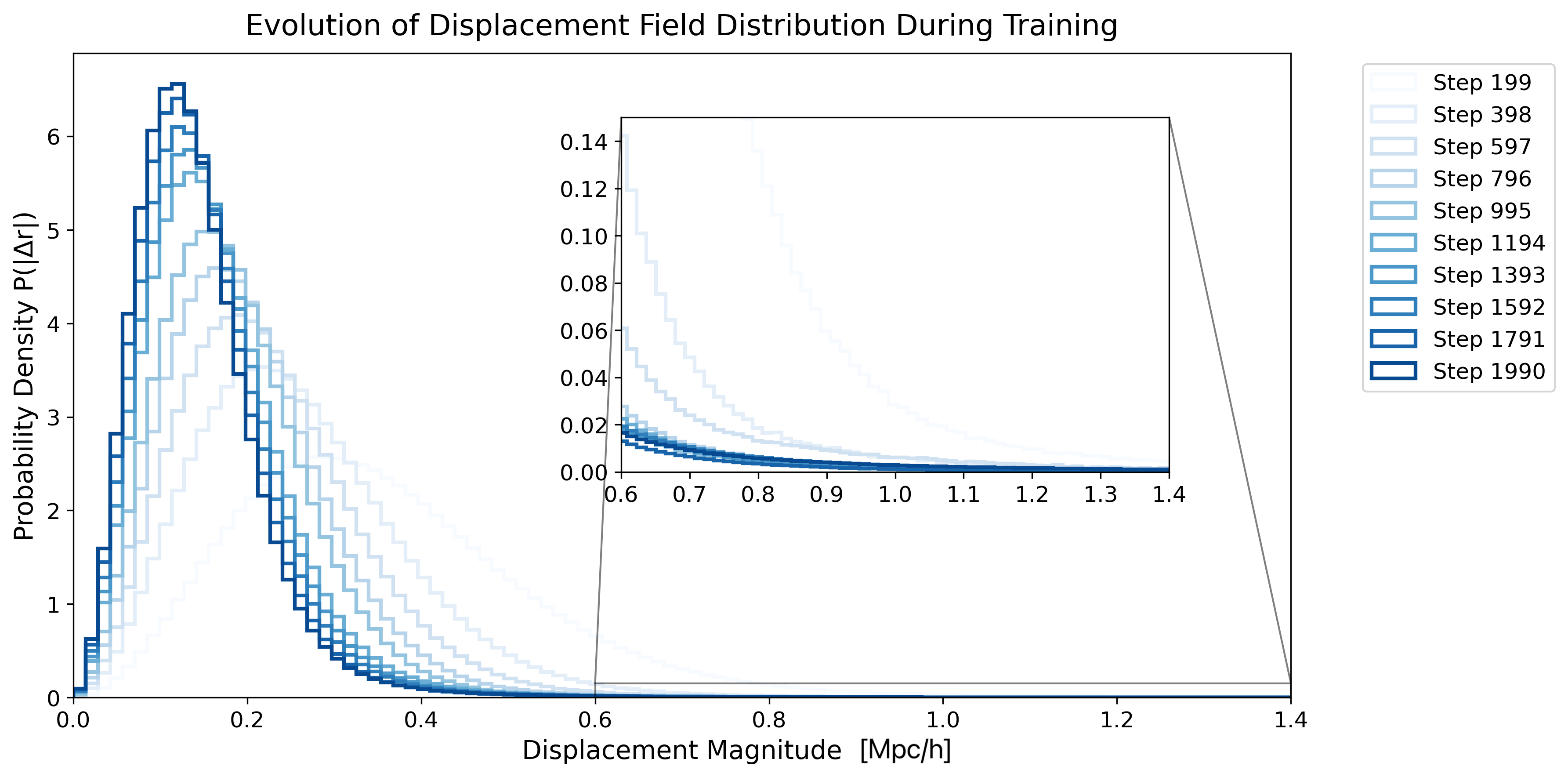}
    \caption{Histograms of the displacement error field contract as the training with the unweighted MSE evolves; with an increase in training steps, displacement errors over every representative particle converge towards lower values. The variance is reduced at the same time, but tails in the error distribution are present (shown in the inset panel). The training dataset is the same as the snapshot pair in Fig.~\ref{fig:displacement_metrics}.}
    \label{fig:errorconvergence-hist}
\end{figure}

To obtain another view of the error distribution, we plot a series of histograms for every displacement error from the field and study how this distribution evolves with training epoch (Fig.~\ref{fig:errorconvergence-hist}). The error histogram peak shifts to lower displacement values and the error variance shrinks as well. While this behavior is expected, nontrivial tails in the error distribution are still manifest (inset panel in Fig.~\ref{fig:errorconvergence-hist}). 

The type of convergence demonstrated in Figs.~\ref{fig:displacement_metrics} and \ref{fig:errorconvergence-hist} is not unexpected since the MSE loss (Eq.~\ref{eq:loss}) is a sum over many points and cannot guarantee uniform local error control. This is one key aspect in which generative mappings of the type considered here differ from numerical PDE solvers, where a {\em local} discretization error is typically estimated and attempted to be controlled. 

Although the type of convergence characterization studied here has value in studying error behavior, it has some of the same drawbacks as global loss functions such as MSE and MAE (Mean Absolute Error) that are characteristic of optimization and machine learning applications. The main issue is that if error properties are dominated by a relatively small number of local domains they may not be sufficiently sampled by the loss function (in contrast to local error metrics) or other averaged quantites, depending on the nature of the averaging.  In Section~\ref{quant}, we will come back to this point when demonstrating the improvement achieved with a density-weighted custom loss function used for training.

\subsection{Density-weighted custom loss function}
\label{sec:densweight}

The convergence results discussed above reveal a clear pattern: while the \emph{global} mean–squared error decreases smoothly with both training-set size and epoch count (Fig.~\ref{fig:convergence}), the \emph{local} error field $|\mathbf{A}-\mathbf{A}'|$ remains dominated by a small fraction of voxels that correspond to highly overdense, strongly nonlinear structures (Figs.~\ref{fig:displacement_metrics}–\ref{fig:errorconvergence-hist}).  
Specifically, 1) particle–wise displacement errors develop long, slowly-shrinking tails associated with dense filaments and (high-density) halo cores; 2) maps of the error field show that these high-error regions coincide almost perfectly with the peaks of the underlying density field (Figs.~\ref{fig:dens-ZA}-\ref{fig:densPM}); and 3) as shown below in Section~\ref{quant}, the power spectrum of the residual field grows steeply toward large $k$, confirming that most of the remaining mismatch resides on small spatial scales.  

Because these overdense regions -- that occupy a small fraction of the overall volume -- carry a disproportionate share of the nonlinear signal that ultimately feeds into a number of physically relevant statistics and covariance estimates, under-weighting them during optimisation may bias the network toward reproducing easy, low-density volumes at the expense of precisely the structures one cares about the most.  
\begin{figure*}
    \centering
    \includegraphics[width=\linewidth]{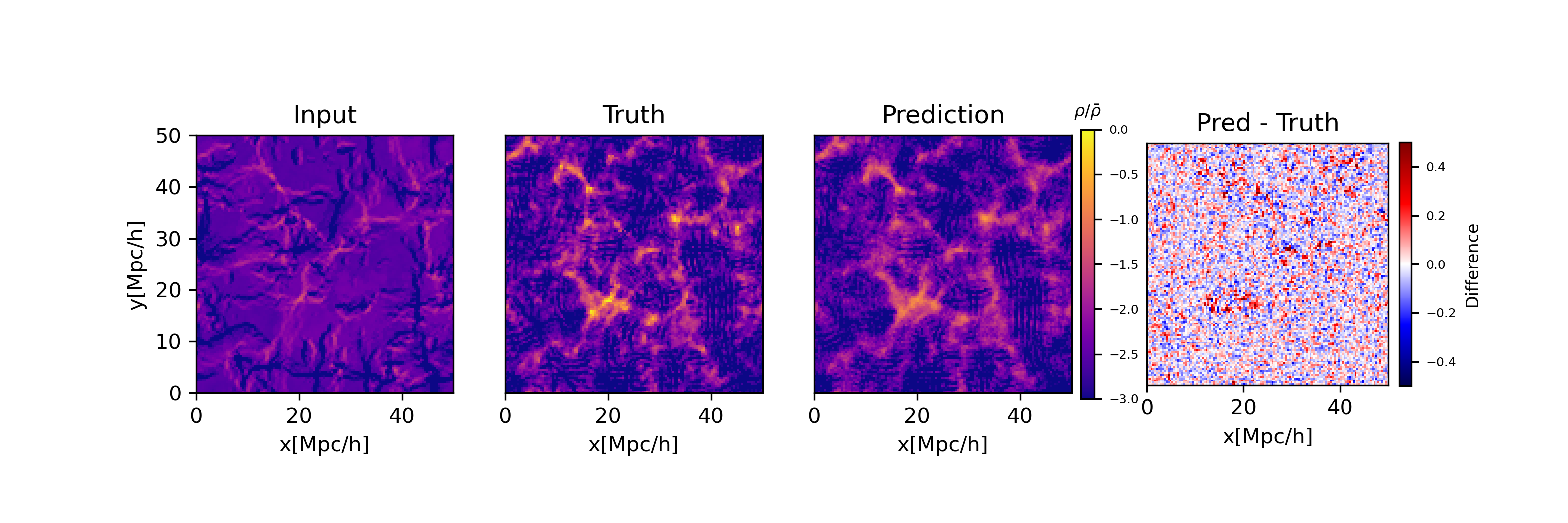}
    \vspace{-1.0em}
    \includegraphics[width=\linewidth]{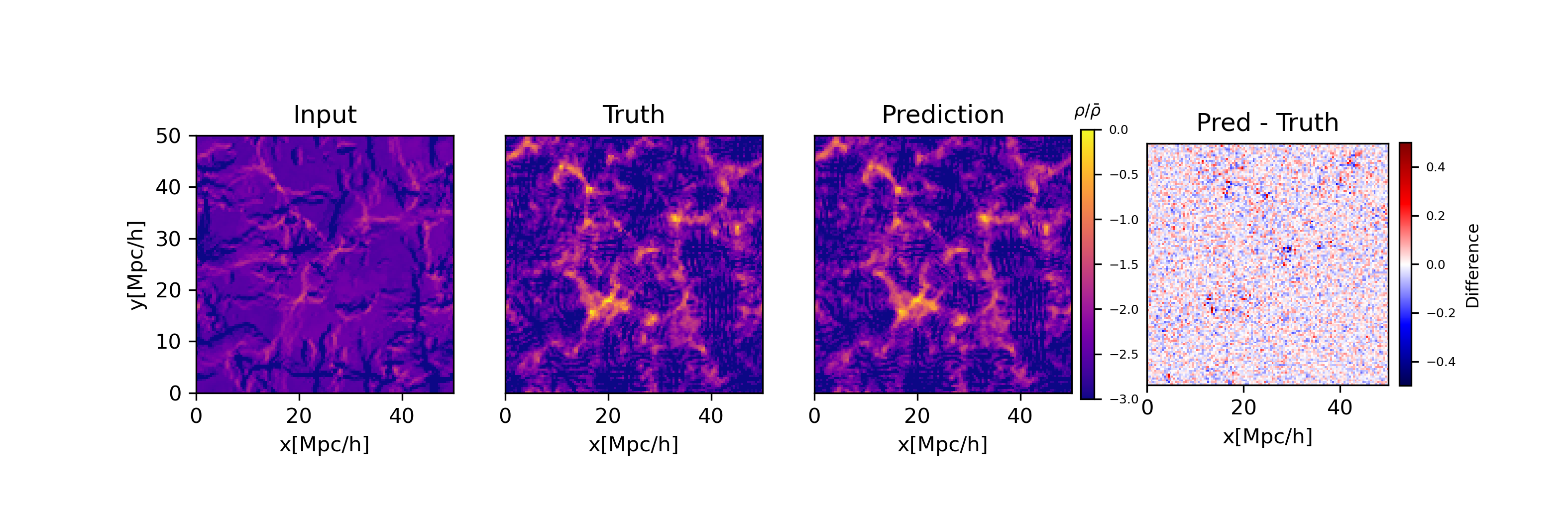}
    \caption{Comparison of the 2-d projected density for the PM simulation test data set ($a=0.05$ to $a=0.2$ snapshots), contrasting the usual MSE loss (upper row) and the density-weighted MSE loss (lower row). As is visually apparent, the small-scale structure is more accurately predicted in the latter case in both spatial resolution and dynamic range. In particular, the middle two panels of the bottom row (``truth'' vs. ``prediction'') are very close and show much improved fidelity compared to the MSE result. The projected error field has smaller excursions and fewer ``hot spots'' in the density-weighted case. }
    \label{fig:density-densityweighted}
\end{figure*}

Motivated by the above arguments, it is natural to introduce a density-weighted loss that would force the model to penalise mistakes in the dense, small-scale regime more heavily, steering the optimisation process toward solutions that are more globally consistent \emph{and} accurate in the physically informative high-density tail.

We therefore implement a density-weighted custom loss function during the training phase which is the usual mean squared error loss, but weighted by the local density at which the local errors are computed. The denser the region is, the more weight is put on the corresponding squared loss contribution:

\begin{equation}
\label{eq:loss_weighted}
    L_{\rm{weighted}} =  \frac{1}{N_p^3}\sum_{i=0}^{N_p} \rho({\mathbf x}^i)\sum_{j=1}^{3} (x^i_{j, \rm{true}}-x^i_{j, \rm{pred}})^2,
\end{equation}
where $\rho({\mathbf{x}}^i)$ is the local particle density. Since we have the input-output pair of particle fields for the training data, the local density can be taken either from the early or the later snapshot. The latter snapshot is where the clustering is higher, so it makes more sense to use that option. Testing both alternatives, we find that this argument is indeed valid as the evaluation metrics show an increased improvement as shown below.

\begin{figure}
   \centering
   \includegraphics[width=\linewidth]{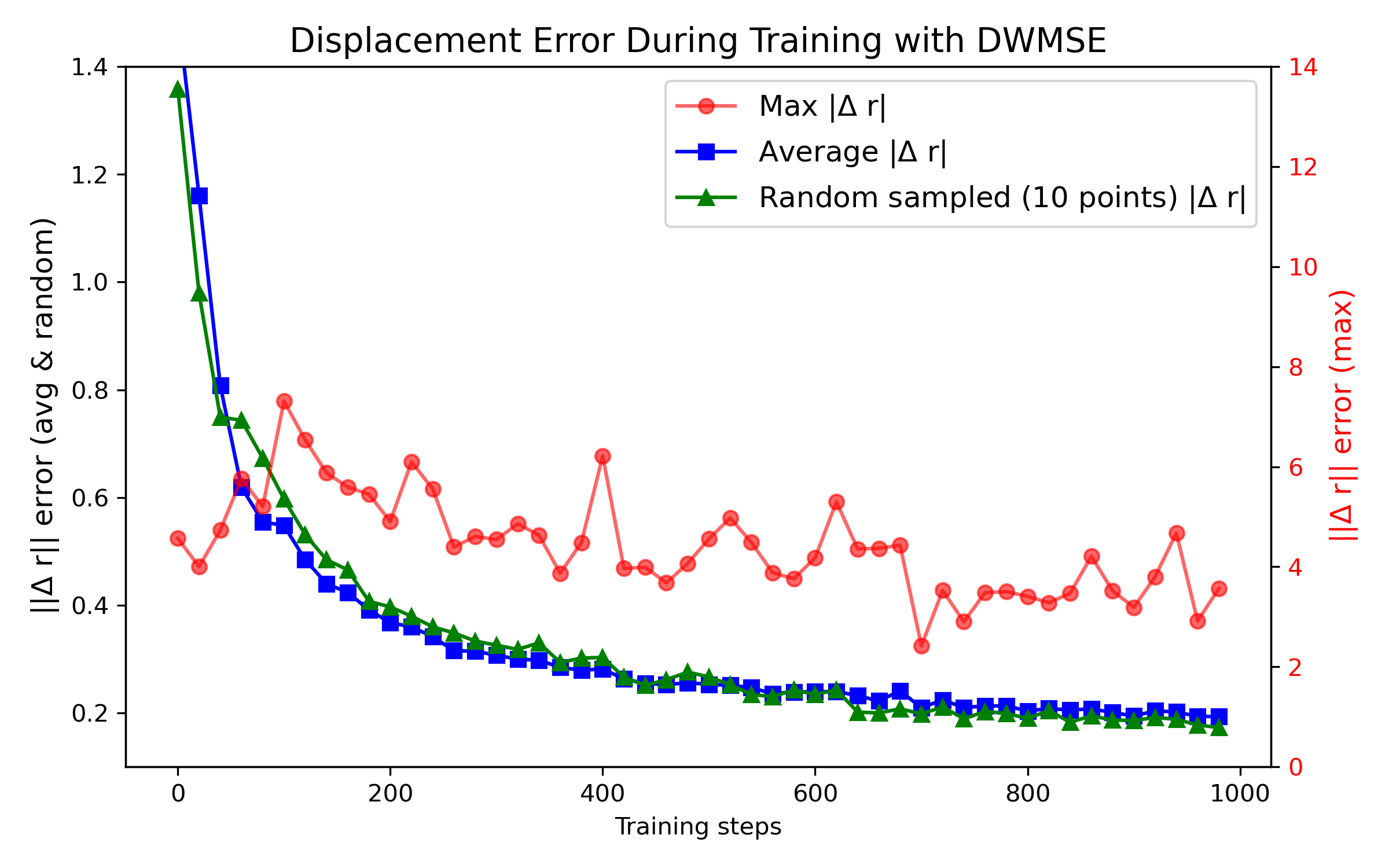}

    \includegraphics[width=\linewidth]{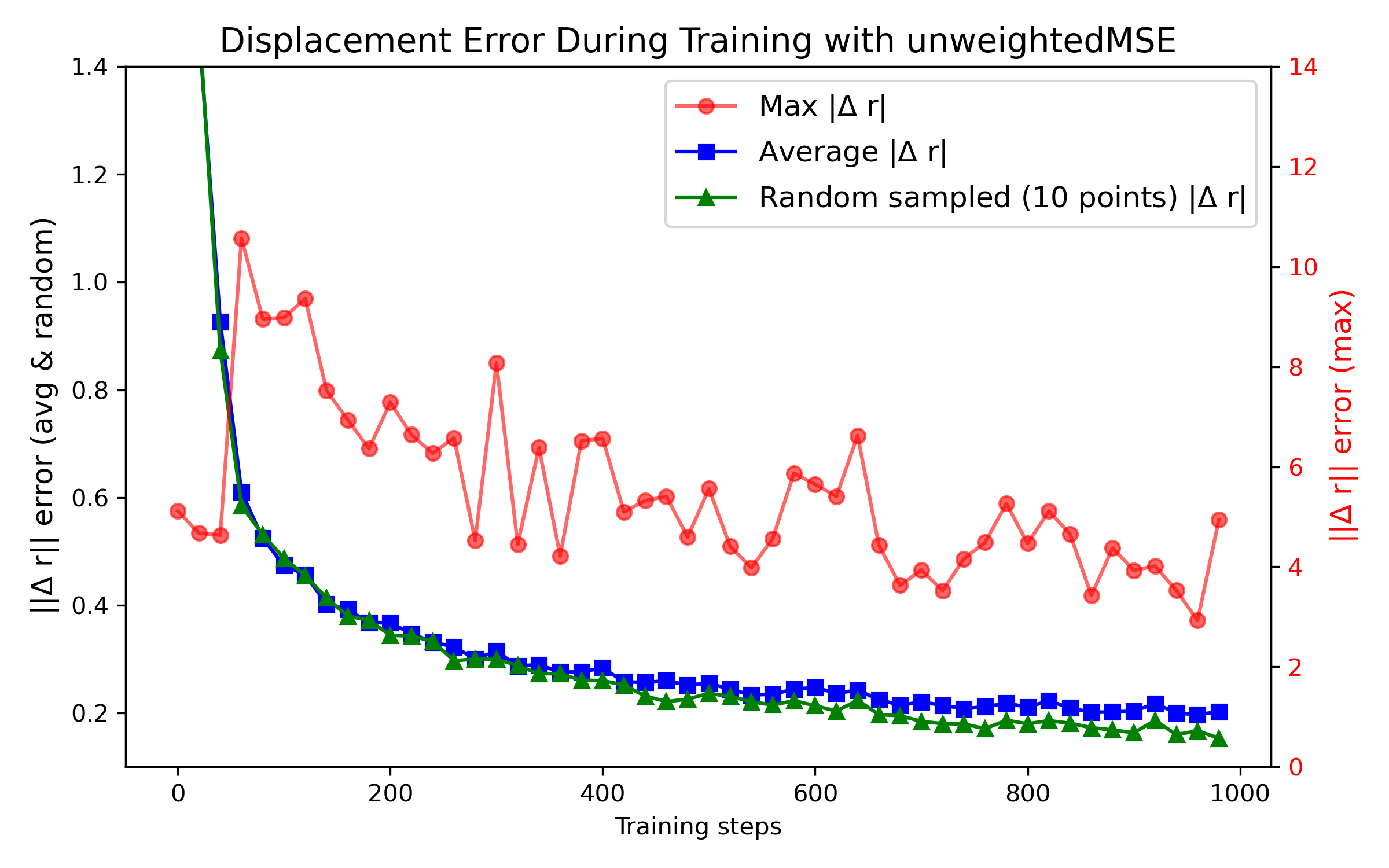}
    \caption{The error convergence behavior of displacement error field for the training with both DWMSE loss (top) and unweighted MSE (bottom), on the $a=0.05$ to $a=0.2$ dataset.}
    \label{fig:displacement_metrics_DWMSE}
\end{figure}
\begin{figure}
    \centering
    \includegraphics[width=\linewidth]{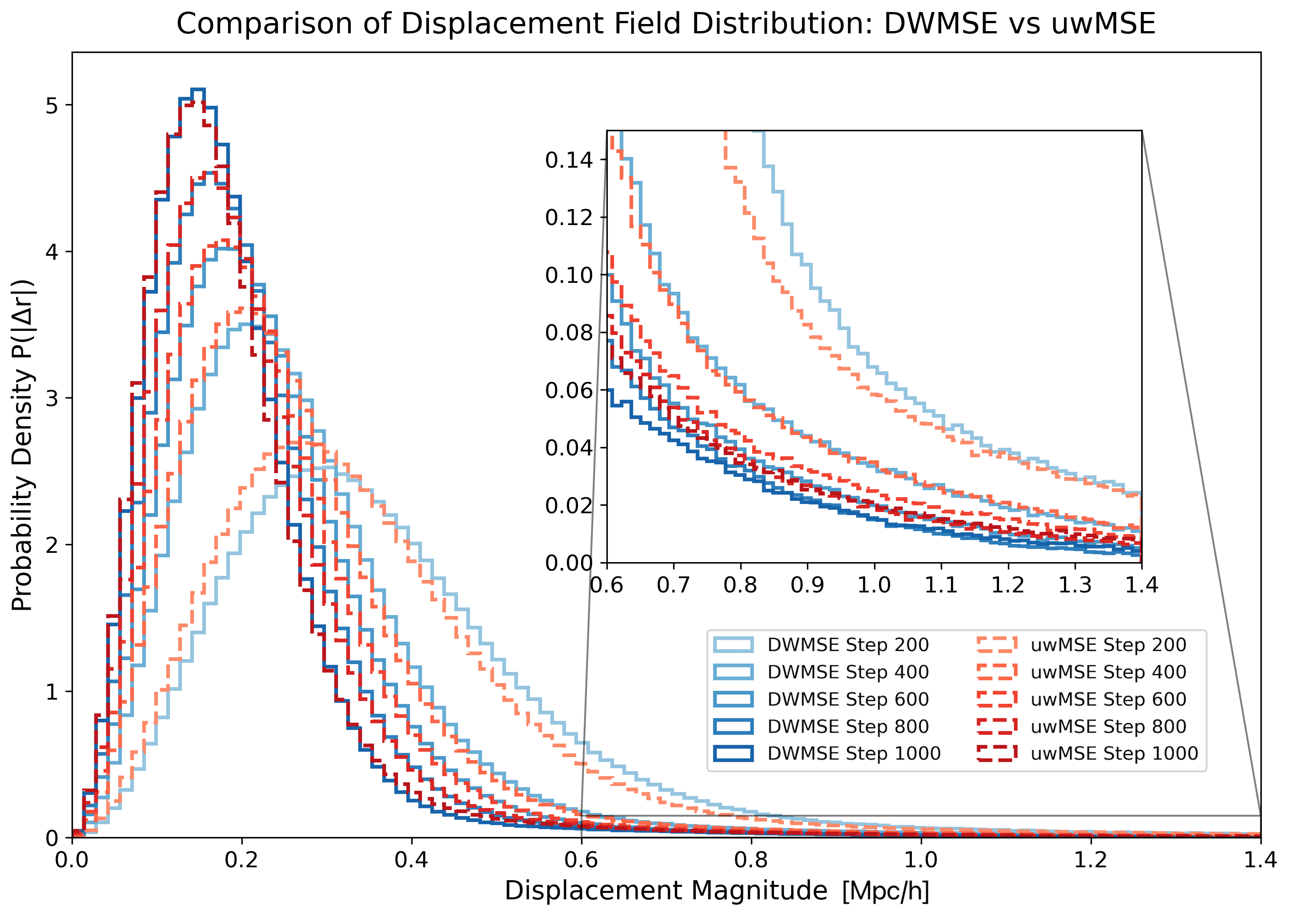}
    \caption{Comparison of the displacement error field histograms for unweighted (red) and weighted (blue) MSE losses for the data set of Fig.~\ref{fig:displacement_metrics_DWMSE}.}
    \label{fig:errorconvergence-hist_dwsme}
\end{figure}
We note that this target density field information is only used during the training process to obtain a more precise field mapping translation, simply assisting the learning process of the network. In the cosmological case, the CIC density estimate-based weighting has a good chance of working well because 1) the data space is low-dimensional (3-d) and 2) the Lagrangian nature of cosmological N-body simulations means that higher density regions are well-sampled with good signal to noise properties.  During the test phase, however, this information is not assumed to be available and the inference (prediction) is still solely dependent on the input displacement field. 

Finally, we note that in principle other weighting functions can be used, including modifications of simple density weighting. We tested higher orders of density weighting, by using $\rho^2$ or $\rho^3$, but results for quantitative metrics such as the power spectrum did not improve as much as compared to the original density weighted case; we leave aside the question of how to optimize the weighting for future work.

\subsubsection{Quantifying improved performance using benchmarks}
\label{quant}

%%%%%%%%%%%%%%%%%%%%%%%%%%
\begin{figure}%[H]
  \centering
 \begin{subfigure}[t]{0.52\textwidth}
 \centering
     \includegraphics[width=\textwidth]{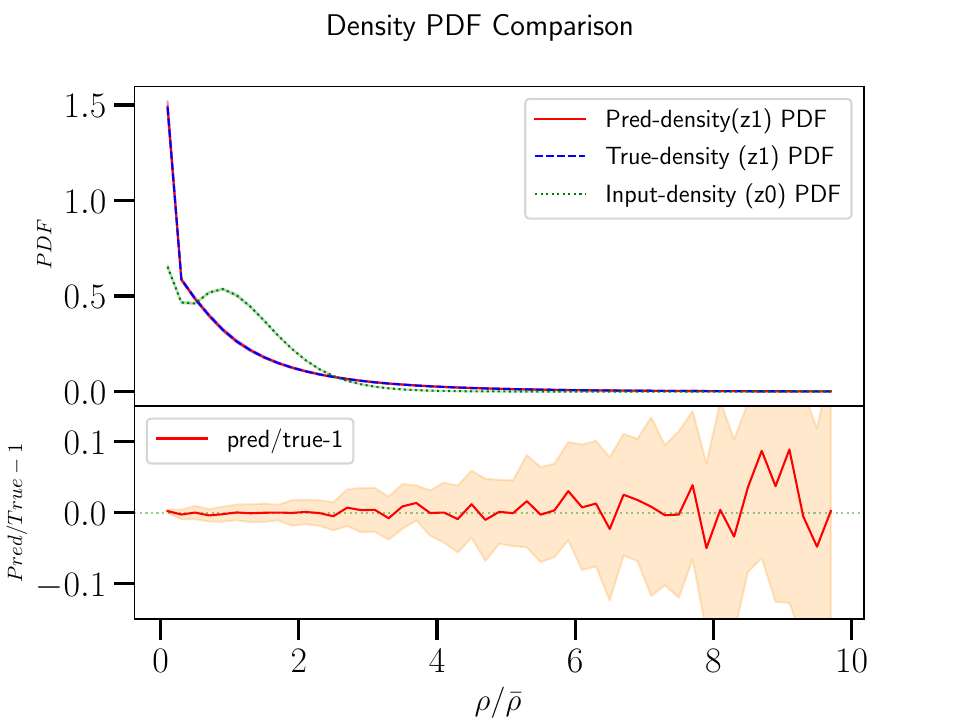}
     % \label{fig:densitypdf}
 \end{subfigure}
\caption{Density PDF curves for the U-Net prediction~($z_1$), true~($z_1$) and input~($z_0$) density fields for the ZA case (see Fig.~\ref{fig:dens-ZA}). The density PDF remains relatively unbiased, although the error variance increases with the density ratio.}
\label{fig:physicalmetrics-PM-pdf}
\end{figure}

\begin{figure}
    \centering
    \includegraphics[width=0.97\linewidth]{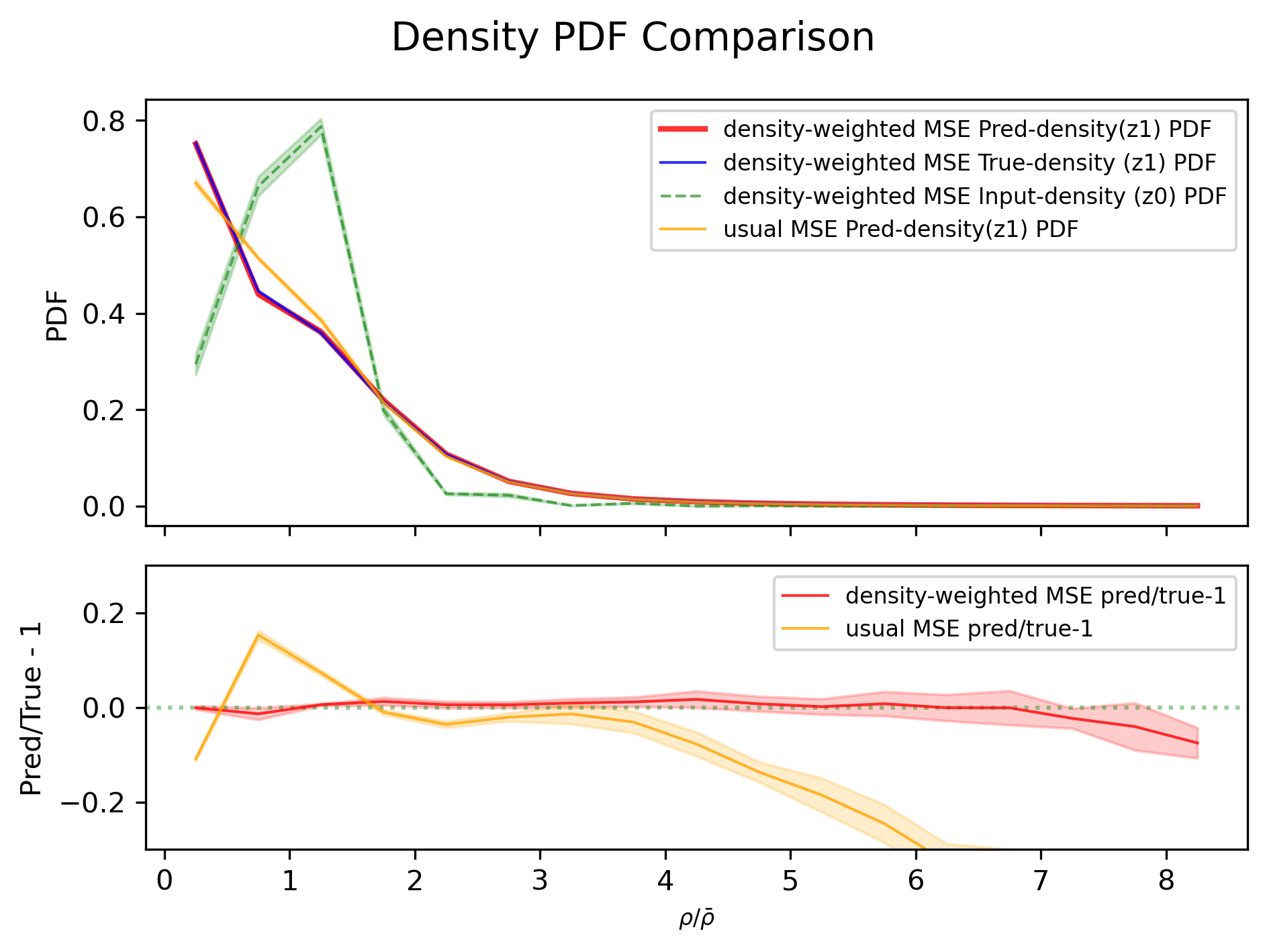}
    \caption{Comparison of the density PDF results on the test set for the unweighted MSE loss and the density-weighted MSE loss. The results using the weighted custom loss function are much improved at all density ratios and significantly extend the density dynamic range (significantly improved behavior at higher densities). }
    \label{fig:dens-PDF-PM}
\end{figure}

%%%%%%%%%%%%%%%%%%%%%%%%%%

%%%%%%%%%%%%%%%%%%%%%%%%%%
%%%%%% P(k)

\begin{figure}
  \centering
  \begin{subfigure}[t]{0.52\textwidth}  
 \centering 
      \includegraphics[width=\textwidth]{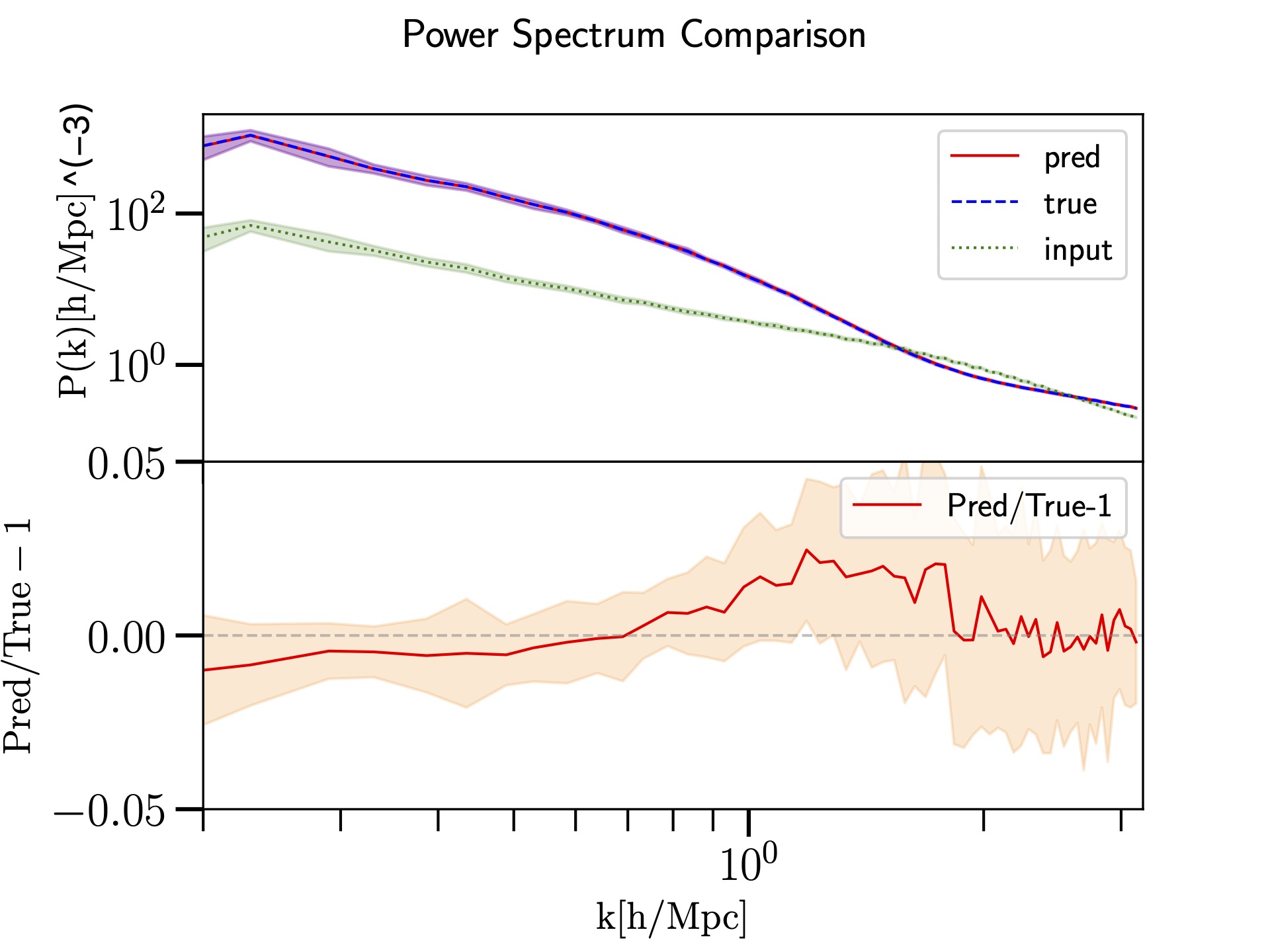}

       \label{fig:pk}
  \end{subfigure}

  \begin{subfigure}[t]{0.52\textwidth}  
 \centering 
      \includegraphics[width=\textwidth]{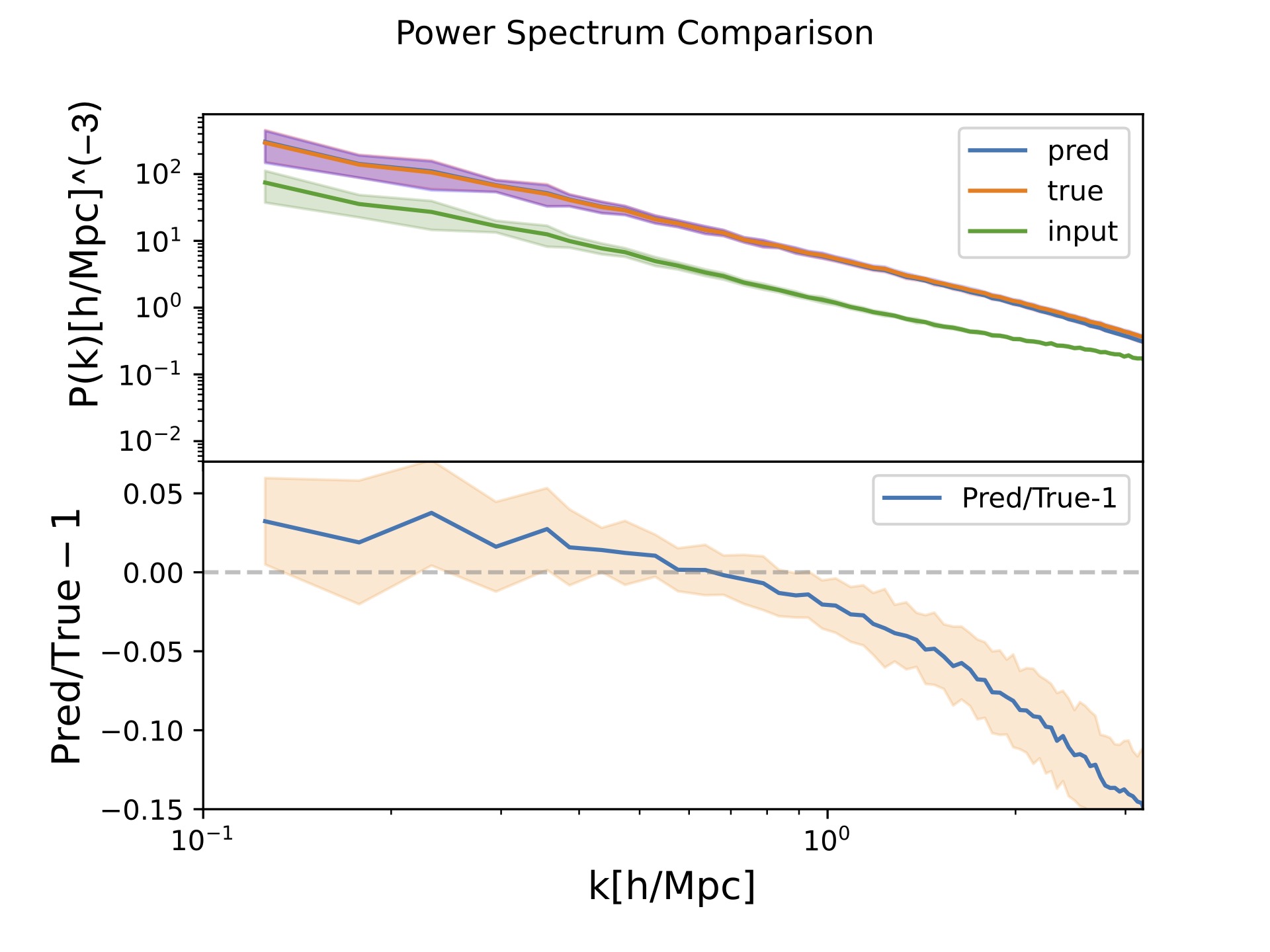}

       \label{fig:pkPM}
  \end{subfigure}
\caption{Matter power spectrum comparison between the predicted density field and ground truth following the conventions of Fig~5. The top panel shows the power spectra of ZA-generated data and training for a=0.046 to a=0.215, while the bottom panel shows the result for PM-generated data for a=0.1 to a=0.2.}
\label{fig:physicalmetrics-Power}
\end{figure}

%%%%%%%%%%%%%%%%%%%%%%%%%%

\begin{figure}
    \centering
    \includegraphics[width=0.97\linewidth]{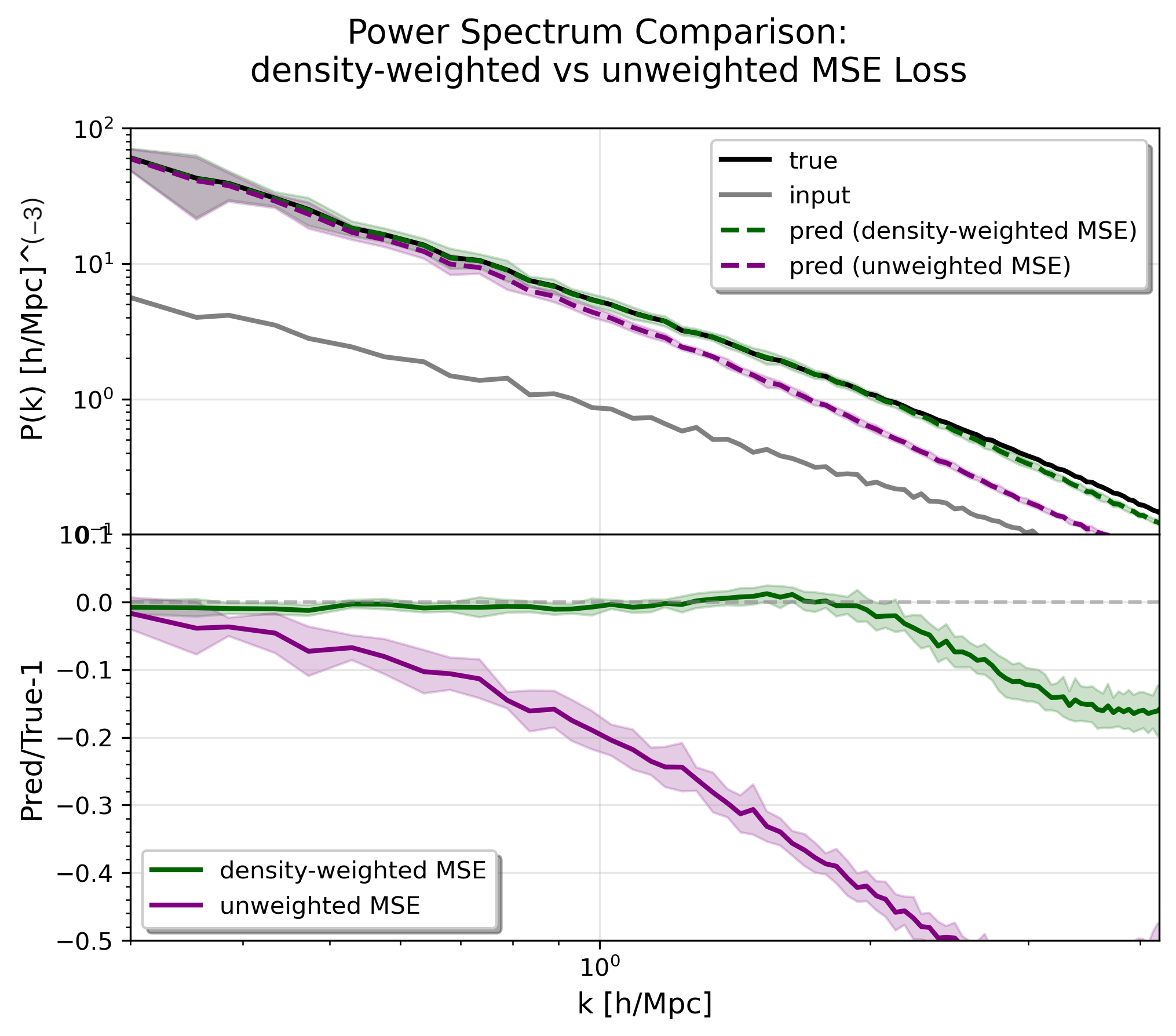}

    \caption{Comparison of the power spectrum results on the test set for comparing the unweighted MSE loss and the density-weighted MSE loss. The weighted custom loss function leads to much improved results over the entire $k$ range and shows excellent agreement with the simulations out to $k=2~h$Mpc$^{-1}$. The dataset tested for this case is PM-generated from a=0.05 to a=0.2. }
    \label{fig:powspec-densityweighted}
\end{figure}

%%%%%%%%%%%%%%%%%%%%%%%%%%

%%%%%%%%%%%%%%%%%%%%%%%%%%%

%%%%%% B(k)

\begin{figure}
  \centering

  \begin{subfigure}[t]{0.52\textwidth}
  \centering
     \includegraphics[width=0.9\textwidth]{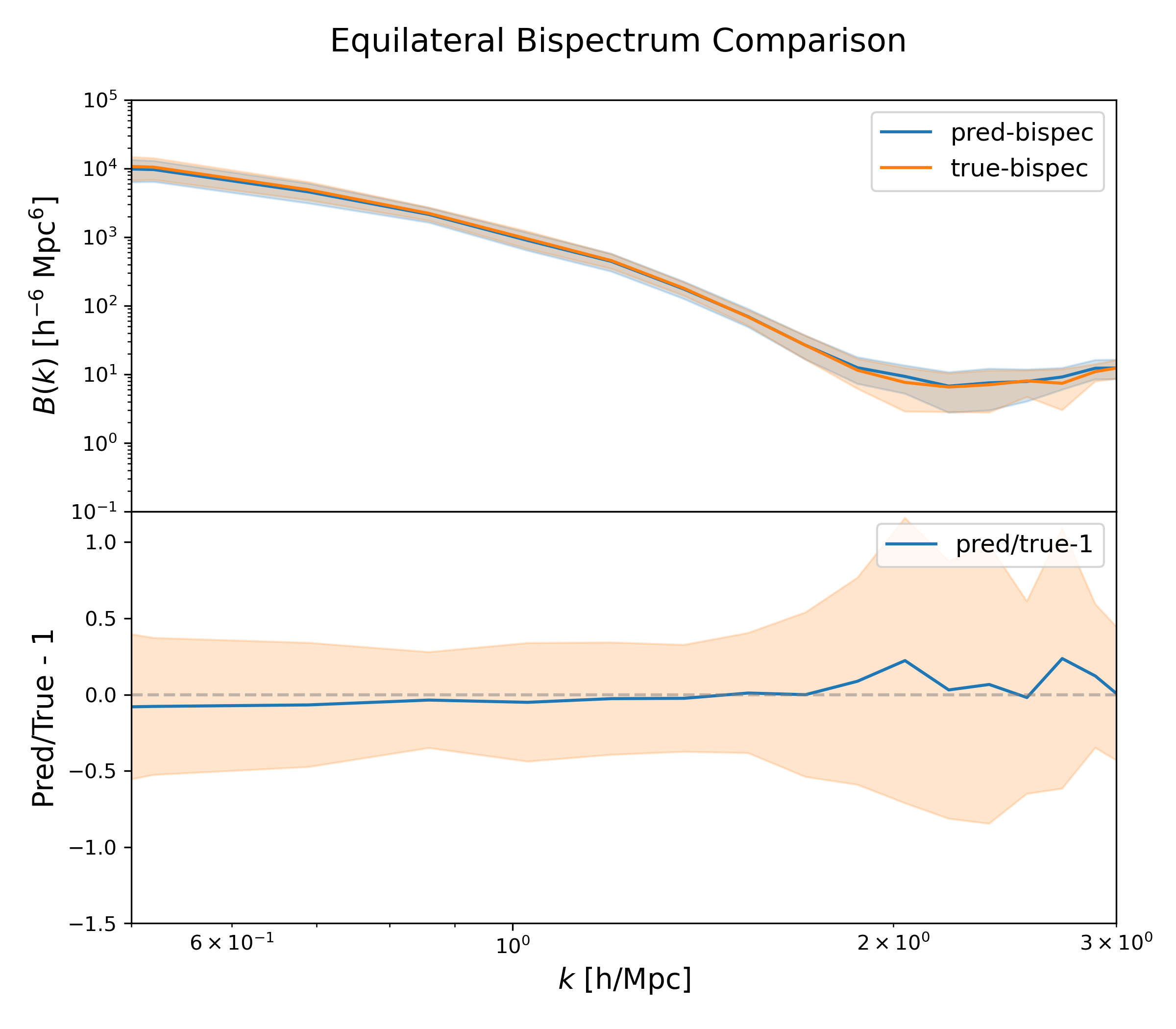}

    \label{fig:bispec}
 \end{subfigure}

  \begin{subfigure}[t]{0.52\textwidth}
  \centering

    \includegraphics[width=0.9\textwidth]{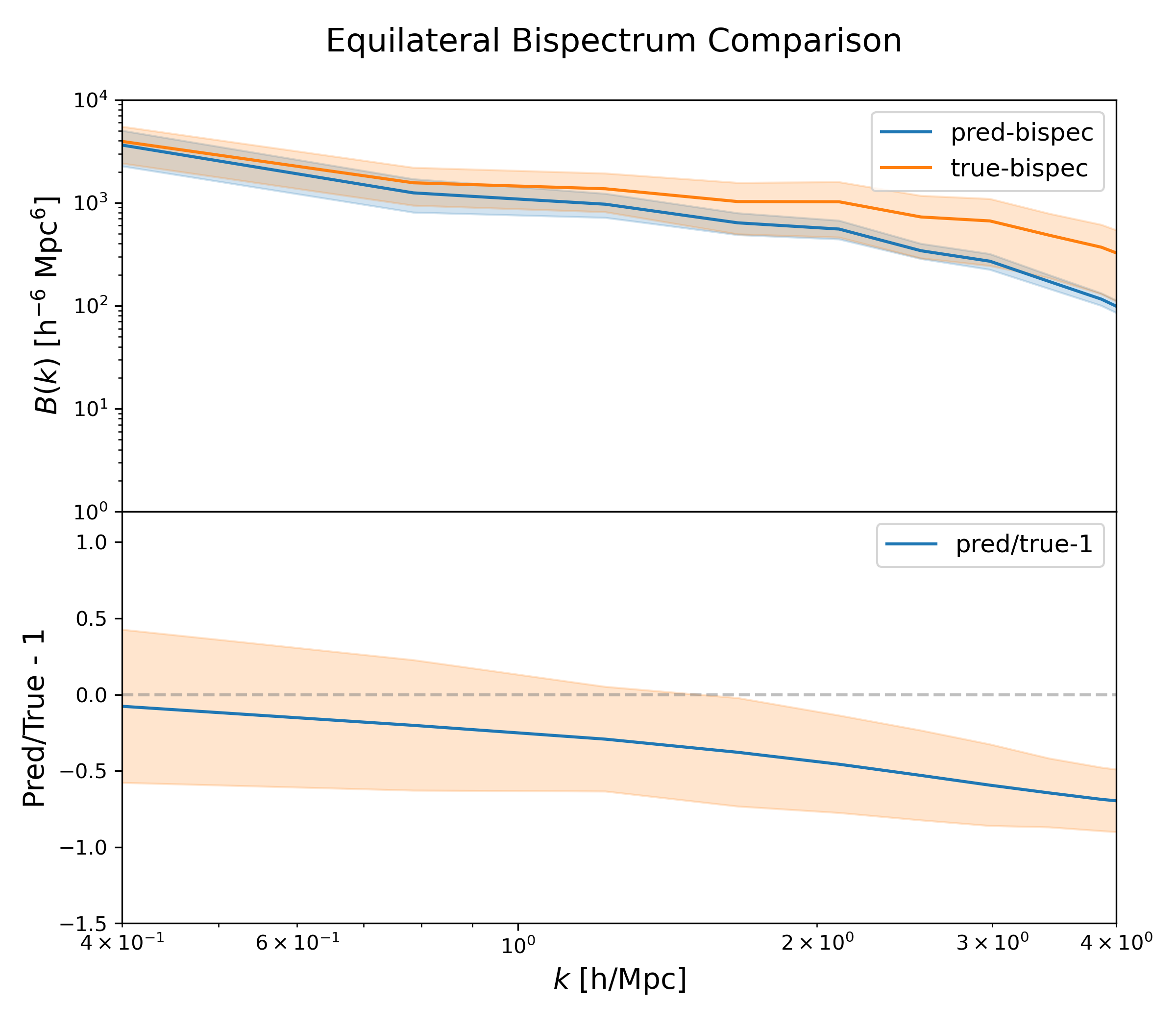}
    \label{fig:bispecPM}
 \end{subfigure}
\caption{The bispectrum comparison between the U-Net prediction and the result from simulations: the top panel shows the results for the ZA evolution for $a=0.215$, while the bottom panel shows the PM-generated results for $a=0.2$. Parameters are for the MSE loss case, the same as for the power spectrum results of Fig.~\ref{fig:physicalmetrics-Power}.}
\label{fig:physicalmetrics-bispec}
\end{figure}

\begin{figure}
    \centering
    \includegraphics[width=\linewidth]{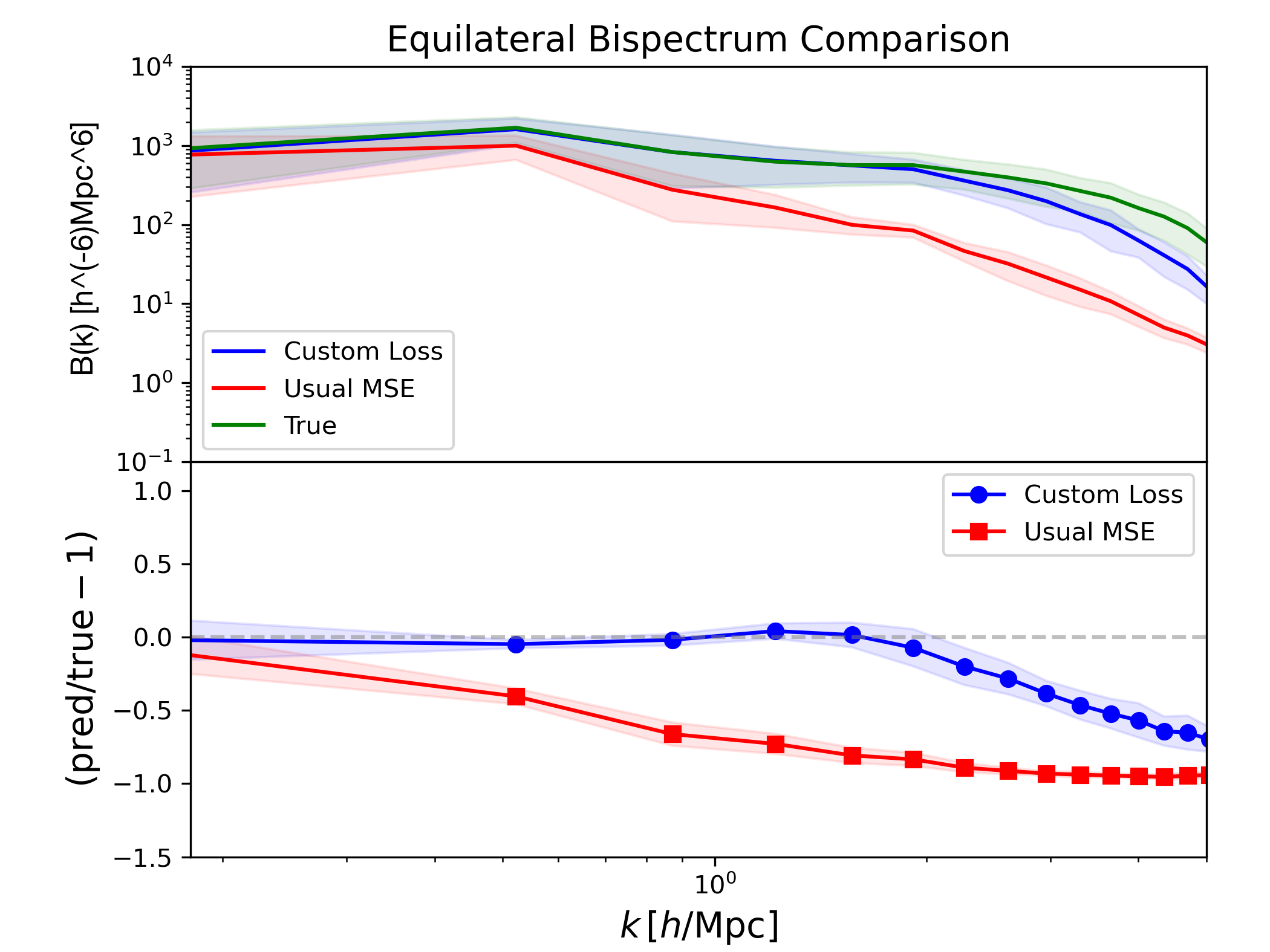}

    \caption{Comparison of the equilateral bispectrum as a function of $k$ from the density field -- prediction versus ground truth -- contrasting density-weighted MSE loss versus the standard MSE loss. Density-weighting improves the agreement on all scales. The dataset tested for this case is PM-generated from a=0.05 to a=0.2. }
    \label{fig:bispec-densityweighted}
\end{figure}

%%%%%%%%%%%%%%%%%%%%%%%%%%

%%%%%%%%%%%%%%%%%%%%%%%%%%
%%%% f1/fES

\begin{figure}
  \centering
 %\hspace*{\fill}%
  \begin{subfigure}[t]{0.51\textwidth}
  \centering
     \includegraphics[width=\textwidth]{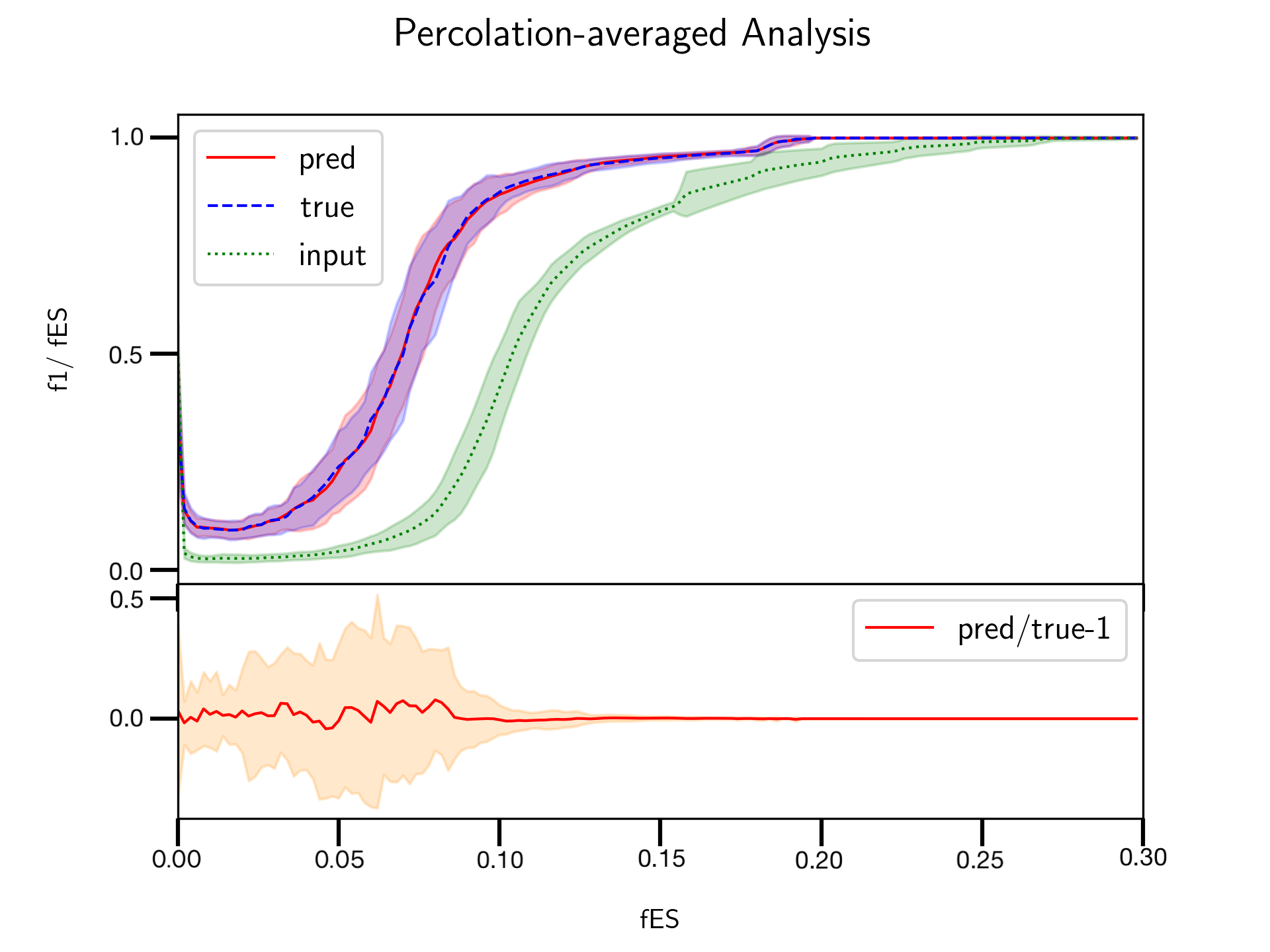}

    \label{fig:percolation}
 \end{subfigure}

  \begin{subfigure}[t]{0.51\textwidth}
  \centering
     \includegraphics[width=\textwidth]{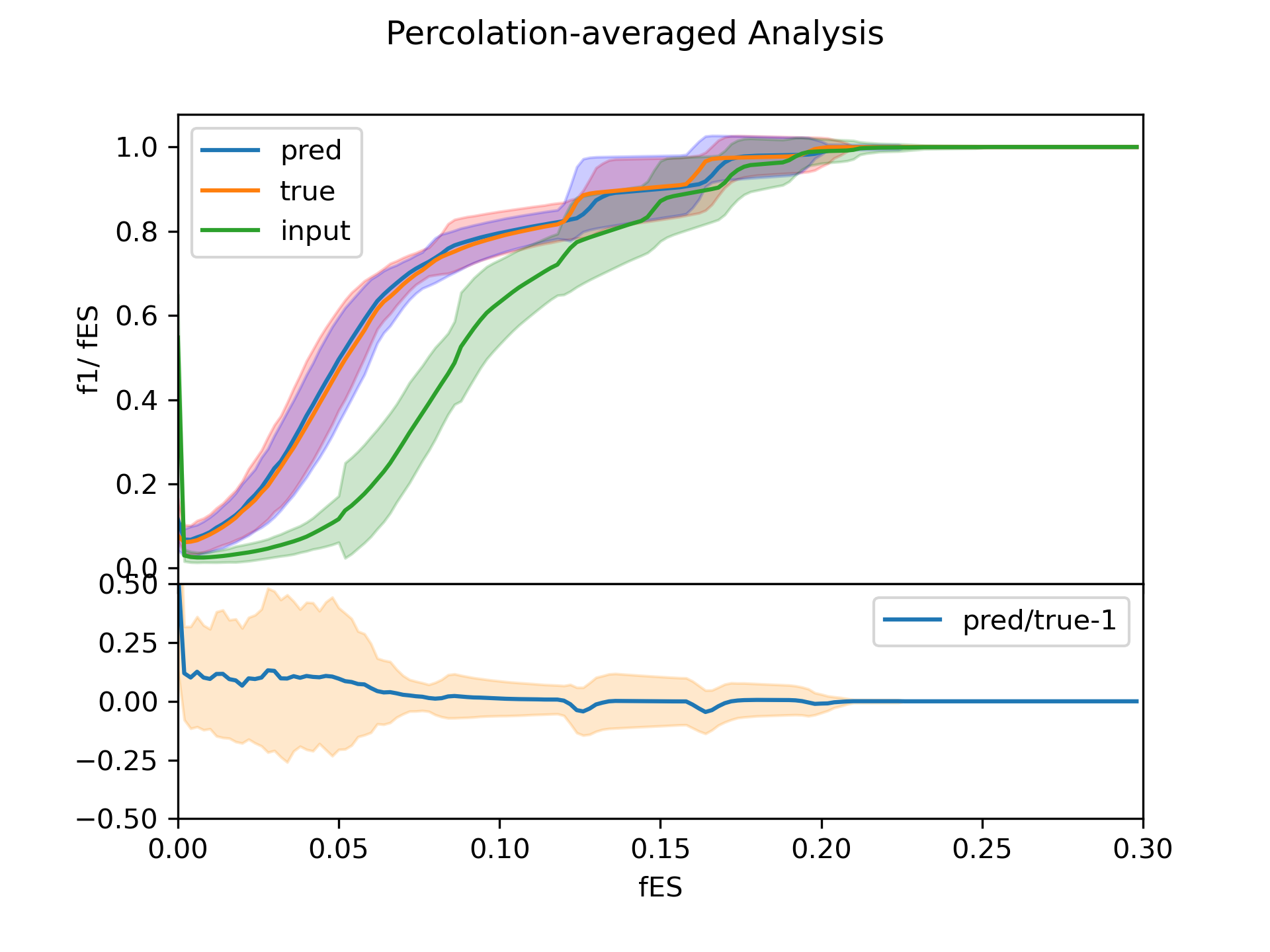}

    \label{fig:percolationPM}

 \end{subfigure}
\caption{Percolation transition analysis as a physical metric for assessment of the U-Net predictions for the ZA (top panel) and PM-evolved (bottom panel) cosmic web.  The filling fraction of the excursion set is $f_{ES}$, which increases as the density threshold is reduced; $f_1$ is the filling fraction of the largest member of the set. The shaded area indicates standard deviations over an ensemble of 30 realizations. The percolation transition is well-captured by U-net; for more discussion see Sec.~\ref{perc2}.}
\label{fig:physicalmetrics-perco}
\end{figure}

\begin{figure}
    \centering
    \includegraphics[width=\linewidth]{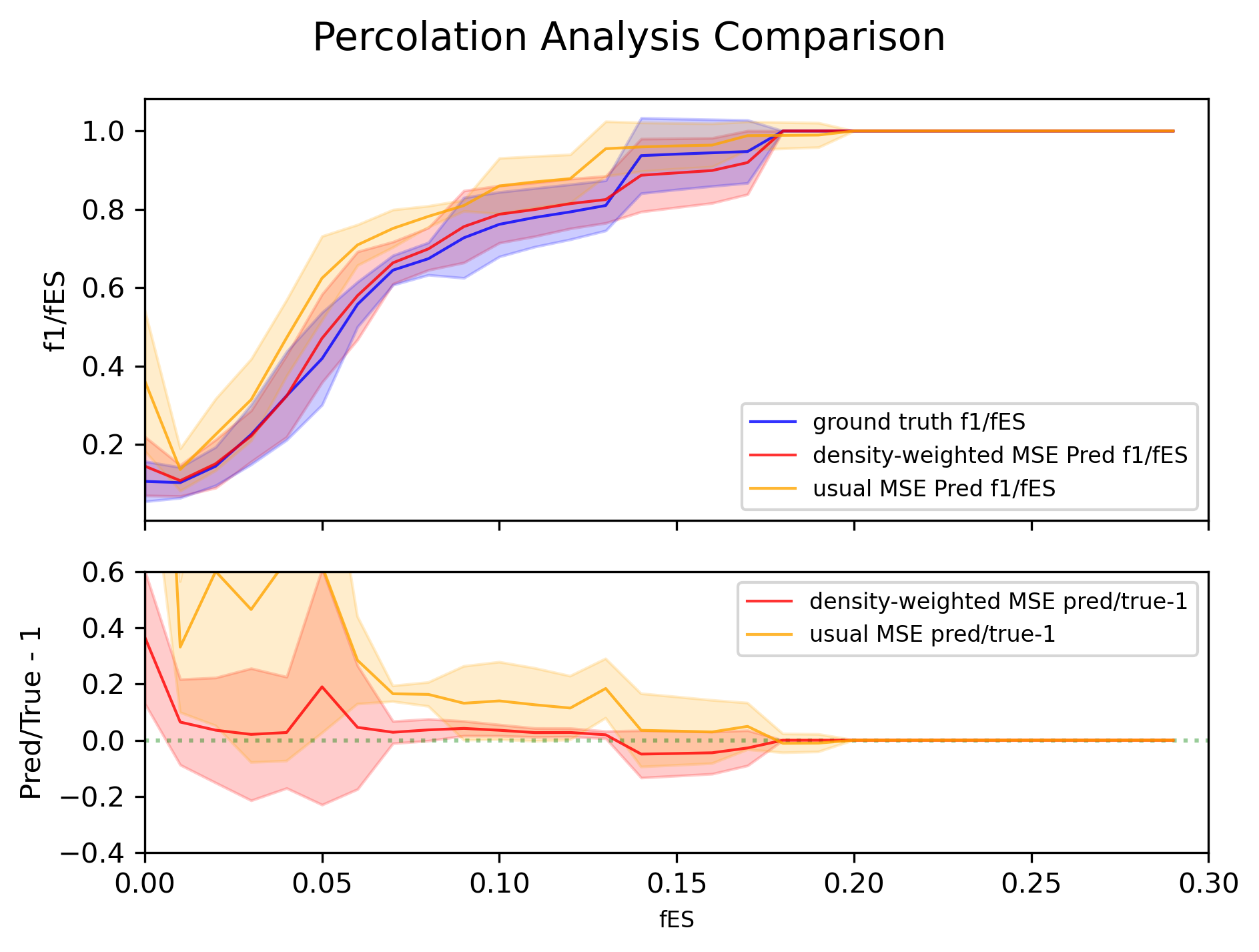}

    \caption{Comparison of the percolation transition contrasting density-weighted MSE loss versus the standard MSE loss averaged on the test set; shaded area indicates standard deviations. Although not as pronounced as with the power spectrum and bispectrum, the density weighted loss results are closer to the numerical data, for further discussion, see Sec.~\ref{perc2}.}
    \label{fig:perco-densityweighted}
\end{figure}

%%%%%%%%%%%%%%%%%%%%%%%%%%

\begin{figure}
  \centering
  \begin{subfigure}[t]{0.49\textwidth}
    \centering
    \includegraphics[width=\textwidth]{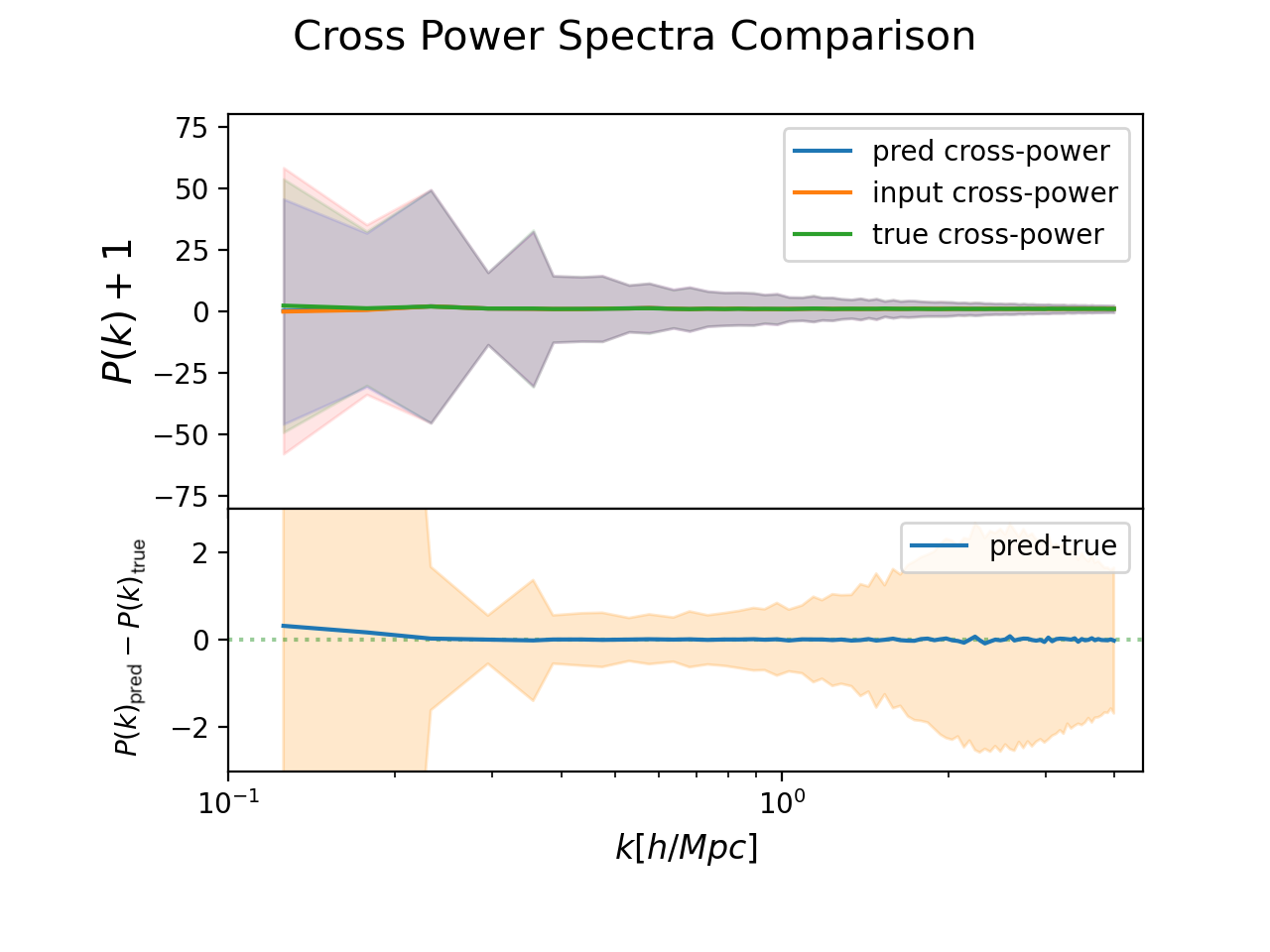}
    \label{fig:crosspow-ZA}
  \end{subfigure}
  \begin{subfigure}[t]{0.49\textwidth}
    \centering
    \includegraphics[width=\textwidth]{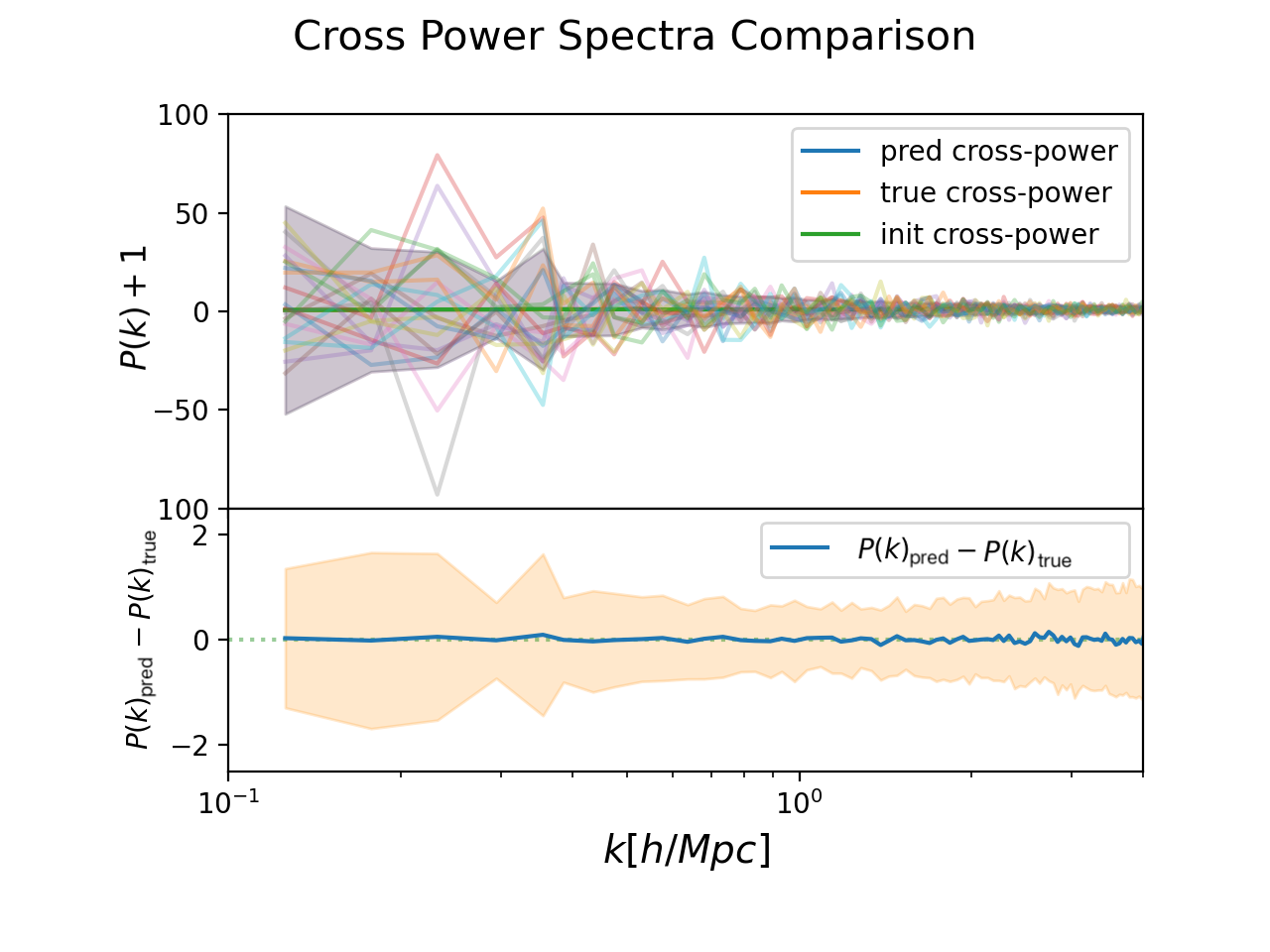}

    \label{fig:crosspow-PM}
  \end{subfigure}
  \caption{Cross-power spectra averaged between different realization pairs from ZA and PM simulations,  respectively. Top panel: Mean and standard deviation of the three cross-power spectra from ZA runs (normalized by the auto-power spectra). Bottom panel: Mean and standard deviation of the cross-power spectra from the PM runs.}
  \label{fig:crosspow-spectra}
\end{figure}

To investigate the potential utility of the density-weighted loss function, we consider an evolution with enhanced nonlinearity and a somewhat more difficult learning setup than considered so far. Evaluations are carried out on a PM dataset evolved over a larger redshift range, from $a=0.05$ to $a=0.2$; 75 training samples in total are used -- a significantly smaller number than were involved in training with the conventional MSE loss since the main purpose here is to carry out a relative analysis. The behavior of absolute errors using the weighted loss function will be investigated separately elsewhere.

To provide a direct comparison, training is carried out for both the usual MSE loss and the density-weighted MSE loss. To provide a first visual impression, the 2-d projected density fields (analog of Fig.~\ref{fig:densPM}) are shown in Fig.~\ref{fig:density-densityweighted}. In these plots, it is immediately apparent that using the density-weighted MSE loss significantly improves both the predicted density field resolution as well as dynamic range at small scales (both the second (truth) and third (prediction) lower panels are strikingly close and a comparison of the upper third panel to the lower one clearly shows the extra smoothing and relative lack of dynamic range for the MSE loss case). Additionally, the projected error field (shown in the rightmost bottom panel) is qualitatively more uniform and has fewer ``hot spots" as compared to the MSE result (corresponding upper panel).

The error distribution fields (Section~\ref{subsec:convg_for_displacement}) in the two cases are compared in Figs.~\ref{fig:displacement_metrics_DWMSE} and \ref{fig:errorconvergence-hist_dwsme} for the $a=0.05$ to $a=0.2$ PM dataset. The quantitative difference between the average errors and the maximum error is not significant (Fig.~\ref{fig:displacement_metrics_DWMSE}) in the later stages of training, although, as shown in Fig.~\ref{fig:errorconvergence-hist_dwsme}, there is an error tail for the unweighted case that goes out further in displacement magnitude.

We now turn to consider the physically important quantitative metrics described in Section~\ref{sec:validation}, to investigate if they are more sensitive to the choice of loss function. 

\subsubsection{The Density PDF}
\label{sec:denspdf}
We begin with the one-point density PDF defined in Section~\ref{sec:pixel}. From its very definition, this metric should be a direct test of how well the density-weighted MSE loss works in improving the dynamic range of the generative map predictions.  

To develop an intuition for how well U-Net performs for this type of prediction, we first consider the ZA case (see Fig.~\ref{fig:dens-ZA}) because it is a simpler dynamical map to approximate. The corresponding result is shown in Fig.~\ref{fig:physicalmetrics-PM-pdf}. As demonstrated in the figure, the smoothness of the ZA evolution allows it to be well-captured by the generative map using the standard MSE loss. The density PDF in this case is essentially unbiased as a function of increasing density ratio (with respect to the mean density), although the error variance increases with the density ratio. This is potentially due to the fact that there are relatively fewer spatial regions sampling high-density excursions, and this could be improved by increasing the volume of the simulation box. 

The density PDF results for the more nonlinear PM evolution are expected to be different, however, following from the differences observed in the projected density fields as visualized in Fig.~\ref{fig:density-densityweighted}. The results are shown in Fig.~\ref{fig:dens-PDF-PM}, for both standard MSE and density-weighted MSE losses. As the top panel demonstrates, the initial density PDF evolves substantially -- the formation of voids is shown by the increase in the PDF for $\rho/\overline{\rho}<1$ and the development of a tail at $\rho/\overline{\rho}>1$, tracing the formation of nonlinear structure (filaments and halos). Consistent with the intuition from Fig.~\ref{fig:density-densityweighted}, we note that the results from the MSE loss are much worse than those from the weighted MSE loss, even at densities not far from the mean density (lower panel of Fig.~\ref{fig:dens-PDF-PM}). The results with the weighted MSE loss are significantly improved at all densities and are unbiased until reaching densities near the upper end of the investigated dynamic range.

The positive results for the density PDF provide good evidence for how well the density field is being predicted, but in order to study how the spatial clustering properties are reproduced, we need to study the matter power spectrum and other measures of spatial statistics, to which we now proceed.

\subsubsection{Matter Power Spectrum}
\label{p_k_discussion}
As in the previous section, we first consider how well the generative map predicts the matter power spectrum for the ZA case. The result is shown in the top panel of Fig.~\ref{fig:physicalmetrics-Power}. Following the previously discussed behavior for the density PDF, we note that the matter power spectrum is also well-predicted, although there is a small residual bias at the percent level. For the PM case (bottom panel of Fig.~\ref{fig:physicalmetrics-Power}), we note a substantial loss of power on nonlinear scales, dropping down to the $\sim10\%$ level, since this is a much more difficult region to predict, as was already seen in the case of the density PDF. (In the case of ZA, there is little evolution of power in this region of the wave number, $k\sim 1~h\mathrm{Mpc}^{-1}$, as shown in the top panel of Fig.~\ref{fig:physicalmetrics-Power}.)

Moving on to the test data set for the weighted MSE loss case with the PM simulations, the results for both loss choices in this test case are shown in Fig.~\ref{fig:powspec-densityweighted}. In the case of the power spectrum, the improvement is quite dramatic and the relative accuracy of the weighted MSE results, as compared to the numerical data, is excellent out to $k\sim 2~h\mathrm{Mpc}^{-1}$, staying at the $1\%$ level. This can be contrasted to the MSE loss case, where the error increases rapidly as $k$ increases, and is already $20\%$ at $k\sim 2~h\mathrm{Mpc}^{-1}$. The final snapshot scale factor is $a=0.2$ corresponding to $z=4$.

\subsubsection{Bispectrum Comparison}
\label{sec:bispectrum_comp}
Going beyond the power spectrum to the (equilateral) bispectrum, we again first consider the ZA case (top panel of Fig.~\ref{fig:physicalmetrics-bispec}) using the conventional MSE loss. As in the case for the power spectrum, the bispectrum results follow a very similar behavior with errors being well-controlled up to a point ($k\sim 1~h\mathrm{Mpc}^{-1}$) beyond which the variance becomes much larger, which is due to the limited resolution of the simulation, resulting in fewer large-scale triangles in the large $k$ region; the equilateral geometry in question has less phase space than more generic bins.

The situation for the PM case parallels the matter power spectrum results, with a substantial suppression of power starting at $k\sim 1~h\mathrm{Mpc}^{-1}$. The variance of the ZA power spectra at higher $k$ values is higher than the PM runs, due to a coarser grid, as well as the increased nonlinearity in the PM simulation and lower particle density leading to more shot noise.

The bispectrum results for the density-weighted MSE loss are shown in Fig.~\ref{fig:bispec-densityweighted}. As for the power spectrum, the weighted loss leads to a significant improvement with very close agreement with the numerical results out to $k\sim 0.3~h\mathrm{Mpc}^{-1}$; although the performance drops off beyond this point, it remains superior to the standard MSE loss across the entire $k$-range considered.

\subsubsection{Percolation Analysis}
\label{perc2}
We now turn to considering a topological metric by analyzing the percolation transition as described in Section~\ref{sec:perc}. The percolation transition for overdense regions occurs at values of the filling fraction of the excursion set, $f_{ES}$, that systematically become smaller the more nonlinear the field is, i.e., as the redshift decreases. This is clearly seen in Fig.~\ref{fig:physicalmetrics-perco} for both the ZA (top panel) and PM (bottom panel) cases. Thus, while the topology of the cosmic web (as viewed by percolation) is in some sense encoded in the initial conditions, it is amplified by the evolutionary map in a way that cannot be captured in linear theory, which does not change the Gaussian nature of the initial conditions~\citep{Percolation}. Therefore, the percolation transition analysis is another way of probing the fidelity of the nonlinear mapping as approximated by U-Net. 

Interestingly, we find that in both the ZA and PM examples, the generative mapping with the MSE loss produces results that are indistinguishable within statistical error from the numerically obtained curves. This is not entirely unexpected since percolation analyses involve working with smoothed fields (typically on the scale of $\sim \mathrm{Mpc}$) and the small-scale loss of resolution seen in Figs.~\ref{fig:dens-ZA} and \ref{fig:densPM} does not appear to affect percolation statistics. (The Gaussian smoothing scale applied for the percolation analysis here is $R_{smooth}$ = 1.5 grid cells.) 

The percolation analysis for the density-weighted loss case follows the expectation from the power spectrum and bispectrum results discussed above. Because the percolation analysis involves smoothed fields, we do not expect a major change, and this is borne out in the data as presented in Fig.~\ref{fig:perco-densityweighted}. We note that as the training set is smaller in this analysis, the results from the MSE loss are worse than those presented in Fig.~\ref{fig:physicalmetrics-perco}. Overall, the results for the weighted loss are closer to the numerical data, but the improvement, as intuitively expected, is modest. 

\subsubsection{Cross-Power Spectrum Comparison}
\label{cross2}
The cross-power test is not primarily a direct probe of the fidelity of the generative mapping, but rather a type of null test checking as to whether the ``memory'' of training sets is leaking into the predictions of the (approximate) dynamical map. This is relevant since in real cosmological applications, one would be concerned about potential sources of systematic error and how to control them. A particular example is covariance matrix estimation, as described in the next section.

As discussed in Section~\ref{cross}, two independent initial conditions run with the actual equations of motion (either ZA or PM) must remain independent under evolution, whereas one may wonder whether the same is true of the map generated by a neural net trained on a {\em finite} number of examples. This issue can be investigated by computing cross-powers (Section~\ref{cross}) and confirming the null result.

The results of the test are shown in Fig.~\ref{fig:crosspow-spectra} for the ZA and PM runs with the standard MSE loss. As shown in the figure, we see no evidence for any memory effect leaking into the cross-power spectra for the training parameters used here. Since the type of loss will not change the basic nature of the results for this test, we do not consider the density-weighted case separately.

\section{Application: correlation matrix and covariance matrix tests}
\label{sec:covariance}
The extraction of cosmological parameters from the power spectrum $P(k)$ traditionally relies on the assumption of Gaussian random fields; however, gravitational clustering progressively induces non-Gaussianity, invalidating this assumption and resulting in inter-band correlations. Therefore, accurately characterizing the statistical properties of power spectrum estimators necessitates a thorough computation of a full covariance matrix that accounts for these non-Gaussian effects~\citep{scoccimarro1999power}. To understand the statistical properties of power spectrum estimators beyond the Gaussian approximation, a calculation of the power spectrum covariance matrix is required. For instance, non-Gaussian effects become most significant on nonlinear scales, where perturbation theory breaks down. It was shown by \cite{scoccimarro1999power} that the non-Gaussian terms in the covariance matrix become dominant for length scales smaller than those corresponding to the nonlinear scale $k_{nl} \sim 0.2~h$Mpc$^{-1}$ at $z=0$, depending on power spectrum normalization. In such scenarios, the hierarchical model becomes an invalid description of the power spectrum covariance matrix in the nonlinear regime.

In practice, covariance tests require a large number of simulation realizations, often thousands or tens of thousands~\citep{takahashi2009simulations}, and are thus computationally very expensive to conduct. On the other hand, utilizing the trained model of a neural network to mimic the actual evolution might open another way to tackle this problem. If a trained network can predict the evolution result of simulations given a large number of input initial conditions, it can be easily used to generate covariance matrices in a computationally efficient way, since these predictions would be many orders of magnitude faster than actual nonlinear numerical computations. Whether the predicted results can reproduce a similar structure in covariance matrices is therefore important and worth further exploration. 

\begin{figure}
  \centering

  \begin{subfigure}{\linewidth}
    \centering
    \includegraphics[width=0.7\linewidth]{\detokenize{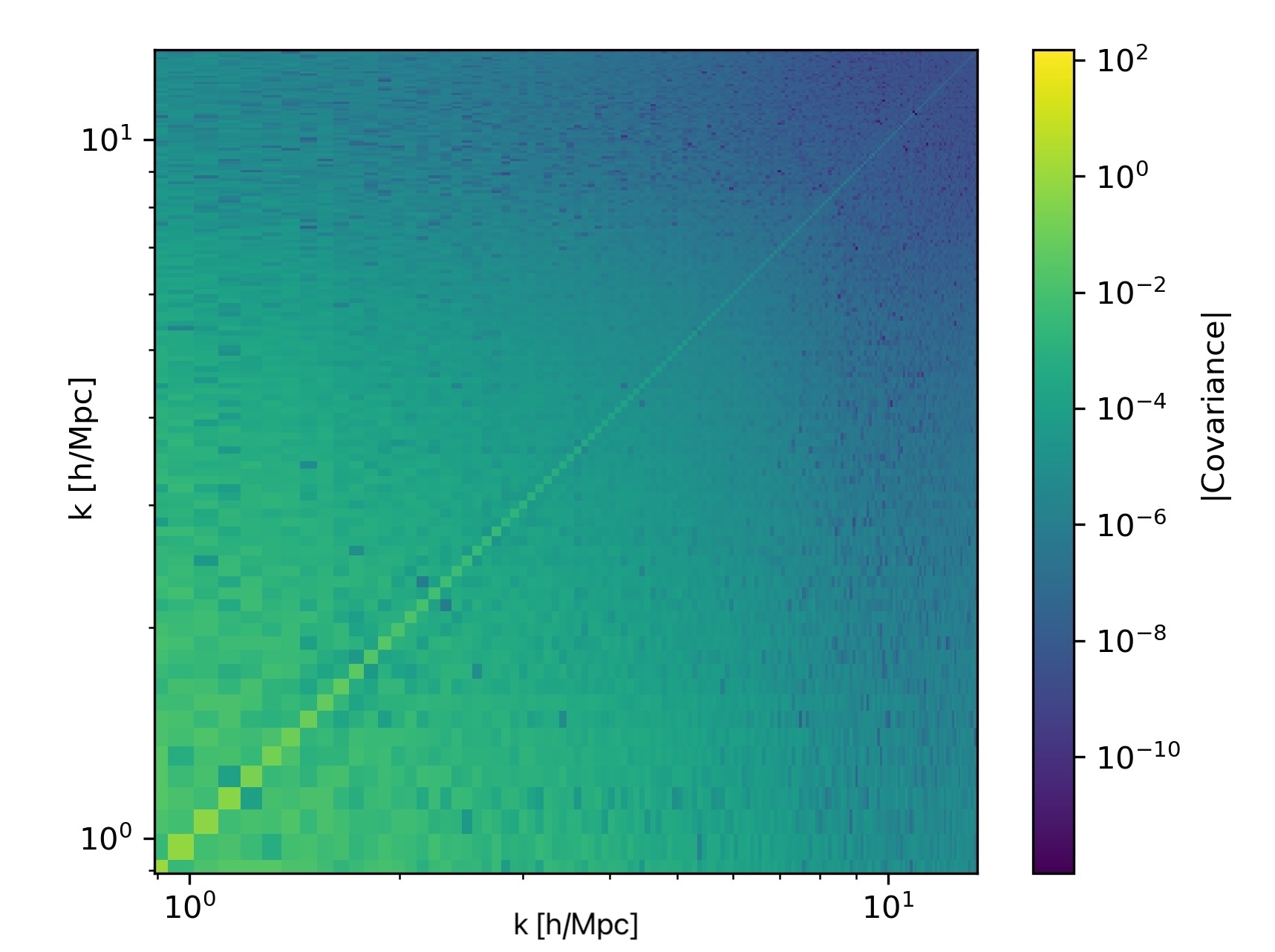}}
    \caption{ZA final snapshot covariance matrix (ground truth).}
  \end{subfigure}%\vspace{-0.8em}

  \begin{subfigure}{\linewidth}
    \centering
    \includegraphics[width=0.7\linewidth]{\detokenize{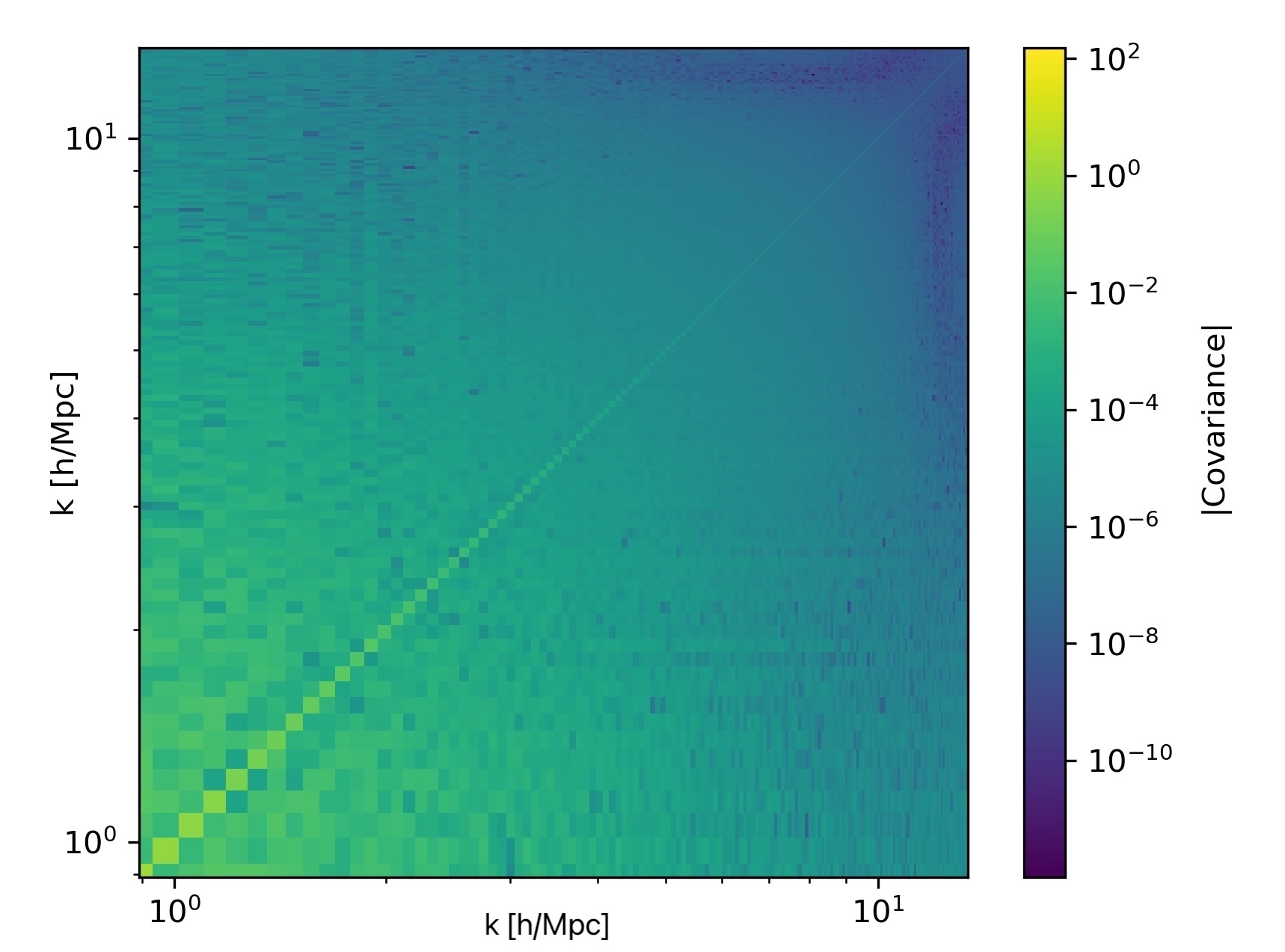}}
    \caption{U-Net (trained on ZA): predicted ZA final snapshot covariance.}
  \end{subfigure}%\vspace{-0.8em}

  \begin{subfigure}{\linewidth}
    \centering
    \includegraphics[width=0.7\linewidth]{\detokenize{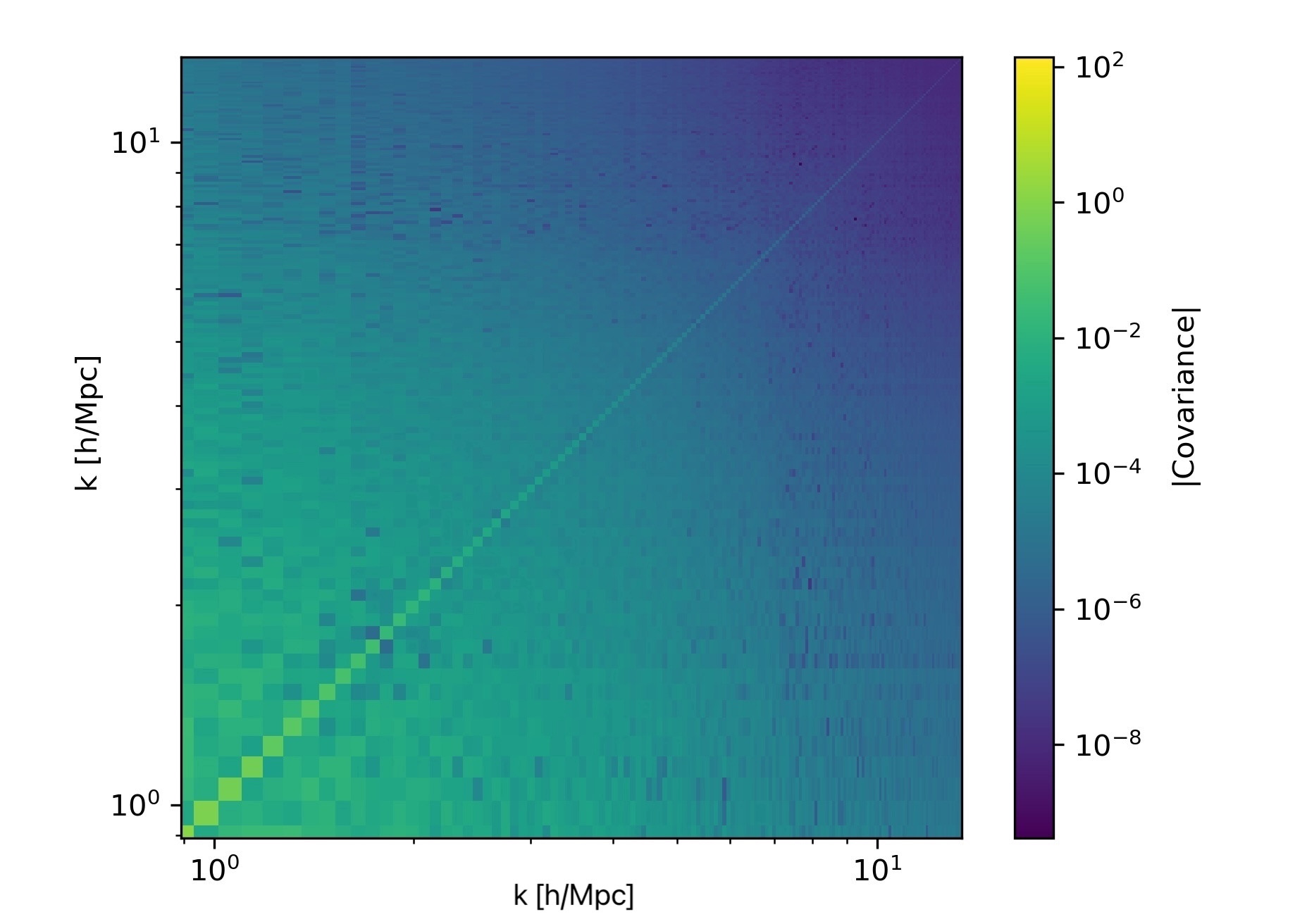}}
    \caption{U-Net (trained on PM, unweighted MSE): predicted PM final snapshot covariance.}
  \end{subfigure}
  %\vspace{-0.8em}
    
  \begin{subfigure}{\linewidth}
    \centering
    \includegraphics[width=0.6\linewidth]{\detokenize{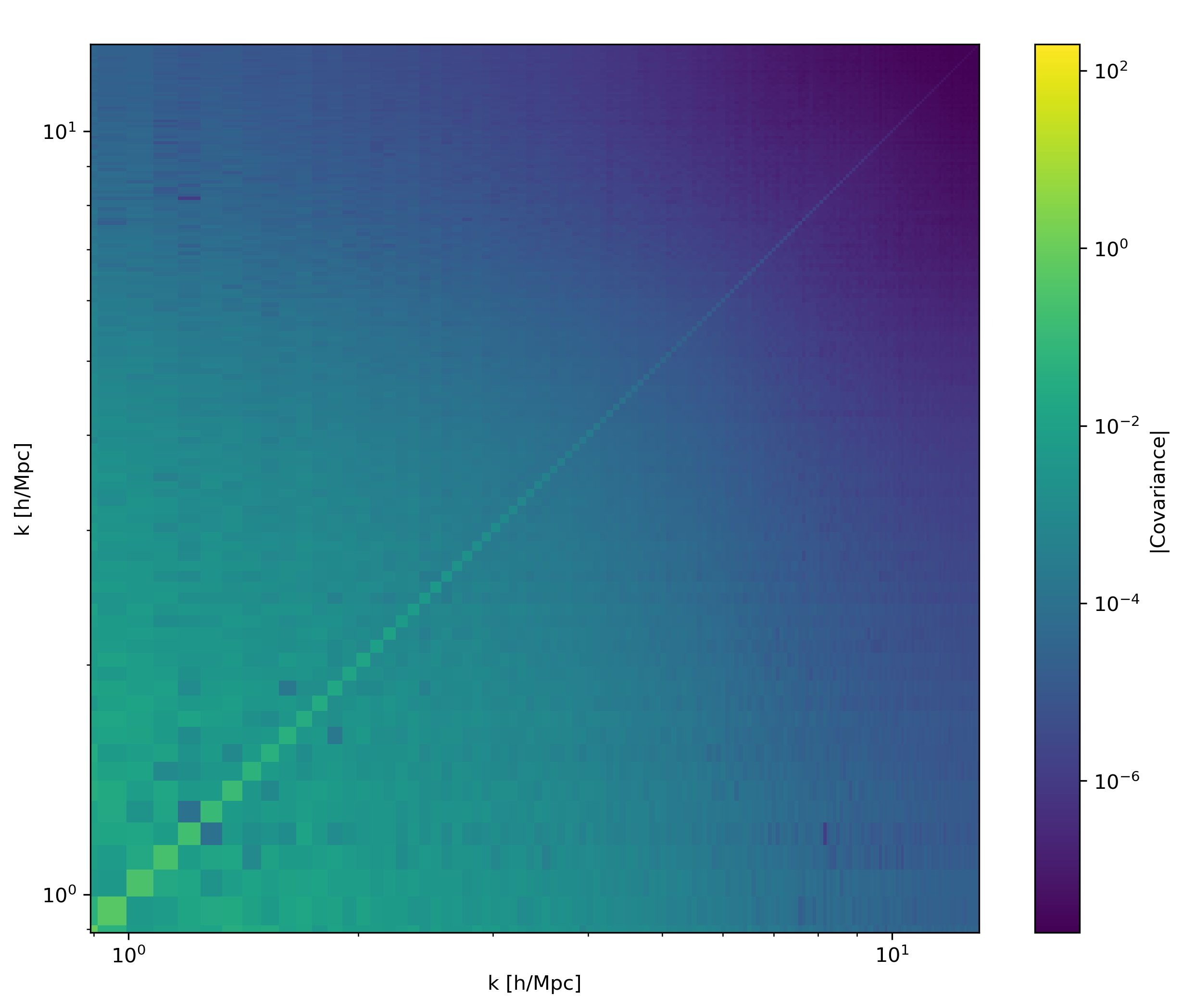}}
    \caption{U-Net (trained on PM, DWMSE): predicted PM final snapshot covariance.}
  \end{subfigure}

  \caption{Covariance matrices $\mathrm{Cov}(k_i,k_j)$ estimated from 1200 realizations from a ZA-evolved dataset and a PM-emulated dataset. From top to bottom: (a) ZA ground truth, (b) U-Net trained on ZA, (c) U-Net trained on PM with unweighted MSE, (d) U-Net trained on PM with DWMSE.}
  \label{fig:cov-cor}
\end{figure}

With our trained deep-learning network, we can test the capability of such a prediction process for covariance calculations. For this test, twelve hundred independent input data samples (at scale factor $a=0.1$) from the test set are input to the  model, which was trained on the mapping between $a=0.1$ and $a=0.2$ with the MSE loss. Analyses of correlation and covariance matrices are then performed on these datasets respectively, for both the ZA and PM runs, as shown in Fig.~\ref{fig:cov-cor}. Comparing the correlation matrix acquired by the true evolution and the predicted evolution for ZA, we find good agreement between the two, despite some differences in the corner regions (i.e., the covariance between small $k$ modes and large $k$ modes). Since we do not have a sufficiently large number of PM runs to test the covariance results -- also the case in reality -- here, we simply compare the U-Net results for the two loss functions, unweighted and density-weighted. The differences between the diagonal covariances are, however, consistent with the error properties of the power spectrum itself when computed using the unweighted and density-weighted loss functions.

\section{Discussion}
\label{sec:discussion}
Deep learning has exhibited a powerful ability in learning complex maps, therefore extending this capability to generating cosmological fields at greatly reduced computational cost as compared to direct simulations is an important direction to pursue. While DL-based translation networks and generative models have demonstrated significant potential, a wide set of critical tests and assessments are necessary before they can be applied in high-precision scientific fields like cosmology (where we are often interested in error metrics aiming at accuracies of better than 1\% \citep{onepercent}). This is the main motivation for the work described in this paper.

Our results presented here underscore the importance of two key aspects: 1) the need for extensive and broad benchmarking capabilities for verification of the generative map approach, and 2) the need for adapting loss functions to the specific fields of interest, here, the nature of the nonlinear cosmic density field.

The importance of comprehensive benchmarking needs to be emphasized. In this work, we have demonstrated the usefulness of a broad class of verification schemes for AI-generated cosmological simulations; this effort is a specific example of motivating the scientific machine-learning community towards using comprehensive sets of benchmarking tools. Starting from a tractable approximation to the complex dynamics of cosmic structure formation, we demonstrate the use of metrics that were not part of the optimization of the neural network. A simple mean-squared error loss does not necessarily enforce or guarantee accuracy for important field-level and clustering statistics. Performance on post-training validation metrics can reveal subtle shortcomings in model predictions. These tests, although performed after the fact, may be incorporated more directly into the training and validation stages to encourage physically consistent mappings.

Standard approaches to improving fidelity place reliance on improving an overall (or integrated) cost function, which can degrade local accuracy as part of an optimization trade-off. Custom losses -- such as the density-weighted one used in this paper -- allow the model to focus on localized denser regions where nonlinear evolution is more pronounced, leading to improved reconstruction of smaller-scale features and partially compensating for the global nature of the cost function. As demonstrated by our results, such approaches can lead to major improvements in capturing the dense filaments and halos that drive most of the nonlinear signal. Future explorations should therefore investigate more sophisticated weighting strategies or hybrid loss functions that incorporate physics-informed constraints.

Our results also emphasize the importance of physics-inspired constraints. While comparing physical benchmarks in detail, as done here, helps to determine the validity of prediction results, the limitation of these testing schemes is that they only occur post-prediction. To make them more useful and lead to more restrictive preservation of physical inputs and laws in the training of AI deep network models, we need to have these physical metrics play a significant role in the model training and validation assessments as well.  Further extending our approach, it would be helpful to investigate how a set of physics-inspired metrics can assist the deep neural network architecture in capturing the physical dynamics and conservation laws underlying the input data. While we have demonstrated the benefits of a custom density-weighted loss, other physical constraints -- such as momentum conservation or invariances in the cosmological fluid evolution -- could be encoded into the architecture or loss functions. Our results also highlight a key consideration regarding costly training datasets: although larger training sets generally yield better agreement with benchmarks, domain-inspired loss functions can substantially reduce errors when data are limited. This is especially important in cosmology, where high-fidelity simulations are often prohibitively expensive.

Meaningful application of DL-generated fields to cosmological studies requires scalability to large enough box sizes and particle numbers; we note that in order for neural networks to compete with simulations, eventually all errors must satisfy demanding constraints (such as being less than $\sim1\%$ for clustering metrics). Additionally, while our initial experiments used relatively small 3-d volumes, practically relevant cosmological applications require simulations with box sizes and dynamic ranges orders of magnitude larger -- the spatial dynamic range of the 3-d NN results must be significantly extended, from a part in a hundred to parts per million. 

Directly scaling up the current approach often exceeds available GPU memory, thus requiring parallelization strategies or multi-scale learning architectures. Moreover, attempts to stitch together smaller boxes (so-called ``collage learning'') highlight the fact that in cosmological applications, boundary conditions and incomplete sampling of large-scale modes can introduce inconsistencies. Future efforts could adopt multi-resolution frameworks, data/model parallelization, or domain-decomposition methods that better respect global modes and periodicity.

Beyond convolutional translators, recent \emph{probabilistic} generative paradigms have shown strong fidelity on scientific data: score-based \emph{diffusion} models (not to be confused with the manifold-learning method “diffusion maps”) learn a noise-to-data denoising process and enable controllable sampling \citep{Song2021ScoreSDE}, while the newly proposed \emph{flow matching}/rectified-flow family unifies diffusion and normalizing flows by directly regressing the continuous-time transport field, often improving sample quality and training stability \citep{Lipman2023FlowMatching}. These directions are complementary to our U-Net–style supervised map and could be adapted to impose physics-aware objectives (e.g., Fourier-space or conservation–aware noise/velocity targets).

Overall, this study demonstrates that although deep learning models --exemplified here by U-Net -- can reproduce large-scale features and pass several validation tests, important discrepancies appear at smaller or more nonlinear scales. Nevertheless, the method proves instructive in identifying key strengths and weaknesses of data-driven approaches to approximating cosmic evolution. We hope that our results, metrics, and recommendations will guide the development of more robust and accurate AI-based cosmological emulators, thereby contributing to next-generation cosmological analyses and surveys. A long and interesting road lies ahead.

\section*{Acknowledgements}

Work at Argonne National Laboratory was supported by the U.S. Department of Energy, Office of High Energy Physics. Argonne, a U.S. Department of Energy Office of Science Laboratory, is operated by UChicago Argonne LLC under contract no. DE-AC02-06CH11357. This material is also based upon work supported by the U.S. Department of Energy, Office of Science, Office of Advanced
Scientific Computing Research and Office of High Energy Physics,
Scientific Discovery through Advanced Computing (SciDAC) program.

The authors would like to thank Ben Gutierrez-Garcia, Michael Buehlmann, and Sandeep Madireddy for useful discussions. The training was conducted on Swing, a GPU system located at the Laboratory Computing Resource Center (LCRC) of Argonne National Laboratory. The analyses performed in this paper utilize the following:  \verb|Numpy| \citep{numpyscipy}, \verb|Matplotlib| \citep{matplotlib}, \verb|Scikit-Learn| \citep{scikit} and \verb|PyTorch| \citep{pytorch}.

\section*{Data Availability Statement}
The data and analysis products, as well as the analysis codes, will be made available upon reasonable request after the acceptance of the paper.

%%%%%%%%%%%%%%%%%%%%%%%%%%%%%%%%%%%%%%%%%%%%%%%%%%

%%%%%%%%%%%%%%%%%%%% REFERENCES %%%%%%%%%%%%%%%%%%

\bibliographystyle{mnras}
\bibliography{mybib}
%%%%%%%%%%%%%%%%% APPENDICES %%%%%%%%%%%%%%%%%%%%%

% \appendix

% \section{Convergence with redshift}
% \input{appendix1}
%%%%%%%%%%%%%%%%%%%%%%%%%%%%%%%%%%%%%%%%%%%%%%%%%%%%%%%%%%%%

\end{document}